
\input phyzzx
\hsize=6truein
\def\TITLEPAGE{\frontpagetrue}
\def\CALT#1{\hbox to\hsize{\tenpoint \baselineskip=12pt
	\hfil\vtop{\hbox{\strut CALT-68-#1}
	\hbox{\strut DOE RESEARCH AND}
	\hbox{\strut DEVELOPMENT REPORT}}}}

\def\CALTECH{\smallskip
	\address{California Institute of Technology, Pasadena, CA 91125}}

\def\AUTHOR#1{\vskip .5in \centerline{#1}}

\def\ABSTRACT#1{\vskip .5in \vfil \centerline{\twelvepoint \bf Abstract}
	#1 \vfil}
\def\ENDTITLEPAGE{\vfil\eject\pageno=1}

\def\sqr#1#2{{\vcenter{\hrule height.#2pt
      \hbox{\vrule width.#2pt height#1pt \kern#1pt
        \vrule width.#2pt}
      \hrule height.#2pt}}}

\def\section#1#2{
\noindent\hbox{\hbox{\bf #1}\hskip 10pt\vtop{\hsize=5in
\baselineskip=12pt \noindent \bf #2 \hfil}\hfil}
\medskip}

\def\underwig#1{	
	\setbox0=\hbox{\rm \strut}
	\hbox to 0pt{$#1$\hss} \lower \ht0 \hbox{\rm \char'176}}

\def\bunderwig#1{	
	\setbox0=\hbox{\rm \strut}
	\hbox to 1.5pt{$#1$\hss} \lower 12.8pt
	 \hbox{\seventeenrm \char'176}\hbox to 2pt{\hfil}}

\def\MEMO#1#2#3#4#5{
\frontpagetrue
\centerline{\tencp INTEROFFICE MEMORANDUM}
\smallskip
\centerline{\bf CALIFORNIA INSTITUTE OF TECHNOLOGY}
\bigskip
\vtop{\tenpoint
\hbox to\hsize{\strut \hbox to .75in{\caps to:\hfil}\hbox to 3.8in{#1\hfil}
\quad\the\date\hfil}
\hbox to\hsize{\strut \hbox to.75in{\caps from:\hfil}\hbox to 3.5in{#2\hfil}
\hbox{{\caps ext-}#3\qquad{\caps m.c.\quad}#4}\hfil}
\hbox{\hbox to.75in{\caps subject:\hfil}\vtop{\parindent=0pt
\hsize=3.5in #5\hfil}}
\hbox{\strut\hfil}}}
\tolerance=10000
\hfuzz=5pt
\def\eslash{\rlap{/}\epsilon}
\def\parslash{\rlap{/}\partial}
\def\Dslash{/\!\!\!\!D}
\def\Aslash{\rlap{~/}A}
\def\eeslash{\rlap{/}e}

\def\vslash{\rlap{/}v}

\TITLEPAGE
\CALT{1860}         
\bigskip
\titlestyle {Combining Chiral and Heavy Quark Symmetries\foot{Work supported
in part by the U.S. Dept. of Energy under Grant no. DE-FG03-92-ER 40701.}}
\AUTHOR{Mark B. Wise}
\CALTECH
\medskip
\centerline{Lectures given at the CCAST Symposium on}
\centerline{Particle Physics at the Fermi Scale}
\centerline{May 27-June 4, 1993}

\ABSTRACT{The chiral and heavy quark symmetries of QCD are reviewed.
These symmetries are used to predict some low--momentum properties of
hadrons containing a single heavy quark.}

\ENDTITLEPAGE

\eject

\noindent {\bf 1. Introduction}

The low--momentum properties of the strong interactions cannot be
described using perturbation theory in the strong coupling constant.
However, there are systematic methods that have proven very useful in
this nonperturbative regime.

In the limit where the light up, down and strange quark masses go to
zero,
QCD possesses an $SU(3)_L  \times SU(3)_R$ chiral symmetry that is
spontaneously broken to the $SU(3)_V$ subgroup.  There are eight
Goldstone bosons (associated with the broken symmetry generators), and
their interactions are described by an effective chiral Lagrangian that is
invariant under $SU(3)_L \times SU(3)_R$.  At low momentum the chiral
Lagrangian can be expanded in derivatives and at the lowest order of
this expansion the self interactions of the Goldstone bosons are
described by a single parameter, the pion decay constant.  It is
possible to treat the quark masses as perturbations and include their
effects.  This is a good approximation provided the light quark masses
are small compared with the chiral symmetry breaking scale.  The quark
mass terms transform under $SU(3)_L \times SU(3)_R$ as $(\bar 3_L, 3_R)
+ (3_L, \bar 3_R)$, and including in the chiral Lagrangian terms that
transform this way explicitly breaks the chiral $SU(3)_L \times SU(3)_R$
symmetry, giving the Goldstone bosons $\pi, K$ and $\eta$ small masses.
It is also possible to describe using an effective Lagrangian, the
interactions of the Goldstone bosons with other particles whose mass
does not go to zero in the chiral limit, e.g., the hyperons.$^{[1]}$

Symmetry methods are also useful for describing the low momentum
properties of hadrons containing a single heavy quark.  In this case it
is appropriate to take the limit where the heavy quark mass goes to
infinity with its four-velocity fixed.  In this limit QCD
possesses a heavy quark spin-flavor symmetry.  Heavy quark symmetry has
proven very useful for describing the interactions of a heavy quark with
light quarks and gluons in the kinematic regime where the light degrees of
freedom typically have a momentum that is small compared with the heavy
quark mass.$^{[2,3,4]}$  For example, heavy quark symmetry implies that all six
form
factors for $B \rightarrow De\bar\nu_e$ and $B \rightarrow D^*e
\bar\nu_e$ semileptonic decays are described by a universal function of
velocity transfer and that this universal function is normalized to unity
at zero recoil (where in the rest frame of the $B$ the $D$(or $D^*)$ is also
at rest).

The interactions of hadrons containing a single heavy quark with the
$\pi, K$ and $\eta$ are constrained by both heavy quark and chiral
symmetry.  Recently there has been a large amount of activity devoted to
examining the implications of this combination of symmetries.  This
paper is meant to provide a pedagogical introduction to chiral
perturbation theory for hadrons containing a heavy quark.  The general
principles of chiral perturbation theory for the Goldstone bosons
are developed and applied to a few examples.  Heavy quark symmetry
will also be introduced.  Finally the two symmetries are combined and
some predictions that use both symmetries are made.

\noindent {\bf 2.  Chiral Symmetries of the Strong Interactions}

The part of the Lagrange density for QCD involving the light quark
fields is
$$	{\cal L} = \sum_a \bar q_a i \Dslash q_a - \sum_a \bar q_a (m_q)_{ab}
q_b \,\, . \eqno (2.1)$$
Here $q_a$ are the light quark fields, $q_1 = u, q_2 = d, q_3 = s$, and
$m_q$ is the light quark mass matrix
$$	(m_q)_{ab} = \pmatrix{m_u & 0 & 0 \cr
0 & m_d & 0 \cr
0 & 0 & m_s\cr} \,\, . \eqno (2.2)$$
In eq. (2.1) $D_\mu$ denotes a covariant derivative
$$	D_\mu = \partial_\mu + ig_s A_\mu^A T^A \,\, ,\eqno (2.3)$$
where $g_s$ is the strong coupling, $A_\mu^A$ denotes the color gauge
field, $A = 1, ..., 8$, and $T^A$ is an $SU(3)$ color generator.

To make the symmetries of eq. (2.1) explicit, it is convenient to introduce
left- and right-handed fields
$$	q_{aL} = {1\over 2} (1 - \gamma_5) q_a\,\, , \eqno (2.4a)$$
$$	q_{aR} = {1\over 2} (1 + \gamma_5) q_a\,\, , \eqno (2.4b)$$
and express the Lagrange density in eq. (2.1) in terms of them.  Actually
$q_{aL}$ and $q_{aR}$ are the more fundamental objects.  They transform
amongst themselves under proper Lorentz transformations.  It is the parity
invariance of the strong interactions that makes it convenient to
combine these two-component fields into a single four-component Dirac
spinor field.  In terms of the left-and right-handed fields
$$	{\cal L} = \sum_a \bar q_{aL} i\Dslash q_{aL} + \sum_a \bar
q_{aR} i\Dslash q_{aR}$$
$$	- \sum_{ab} \left[\bar q_{aL} (m_q)_{ab} q_{bR} + \bar q_{aR}
(m_q)_{ab} q_{bL}\right] \,\, . \eqno (2.5)$$

Let's examine eq. (2.5) in the limit $m_q \rightarrow 0$.  This will end up
being a good approximation because the up, down, and strange quark
masses are small compared with the chiral symmetry breaking scale.  In
this limit
$$	{\cal L} = \sum_a \bar q_{aL} i\Dslash q_{aL} + \sum_a \bar
q_{aR} i\Dslash q_{aR} \,\, , \eqno (2.6)$$
and ${\cal L}$ possesses the global chiral symmetry $G = SU(3)_L \times
SU(3)_R$, under which
$$	q_{aL} \rightarrow L_{ab} q_{bL}\,\, , \qquad L\epsilon SU(3)_L  \eqno
(2.7a)$$
and
$$	q_{aR} \rightarrow R_{ab} q_{bR}\,\, , \qquad R\epsilon SU(3)_R\,\, ,
\eqno (2.7b)$$
(the repeated index $b$ is summed over $1,2,3$).

Eqs. (2.7) represent a symmetry of the Lagrange density, eq. (2.6), but
not of the vacuum.  In QCD the quark bilinear
$\bar q_{aR} q_{bL}$ has the vacuum expectation value
$$	< 0 | \bar q_{aR} q_{bL}|0> = v \delta_{ab} \,\, . \eqno (2.8)$$
Performing a $SU(3)_L \times SU(3)_R$ transformation on the vacuum
state, we find
$$	<0| \bar q_{aR} q_{bL}|0> \rightarrow <0|\bar q_{cR} R_{ac}^*
L_{bd} q_{dL} |0>$$
$$	= R_{ca}^{\dag} L_{bd} <0| \bar q_{cR} q_{dL} |0> = R_{ca}^{\dag}
L_{bd} v\delta_{cd}$$
$$	=v (LR^{\dag})_{ba}\,\, . \eqno (2.9)$$
So only if the transformation is in the vector subgroup $SU(3)_V$ where
$L = R$ is the vacuum invariant.  The symmetry $G = SU(3)_L \times
SU(3)_R$ is spontaneously broken to the vector subgroup $H = SU(3)_V$.
Because $G$ is a symmetry, transformations in the coset space $G/H$ take
the vacuum state into another state of the same energy.  Consequently
the theory must contain massless particles, one for each broken
generator of $G$.  To ensure that field configurations related by $G/H$
transformations have the same energy, these massless particles are
derivatively coupled.

Before proceeding further with QCD, let's consider the simpler example of a
theory
with a single complex scalar field $\phi$ and the Lagrange density
$$	{\cal L} = \partial_\mu \phi^* \partial^\mu \phi - \lambda
(|\phi|^2 - v^2)^2 \,\, . \eqno (2.10)$$
The Lagrange density, eq. (2.10), is invariant under the global $U(1)$ symmetry
$$	\phi \rightarrow e^{i\Omega} \phi \,\, . \eqno (2.11)$$
This symmetry $G = U(1)$ is spontaneously broken by the expectation
value
$$	<0|\phi| 0> = v \,\, . \eqno (2.12)$$
The resulting Goldstone boson field can be thought of as arising from
transforming $\phi$ away from its vacuum expectation value by an element
of $G$.  For (low energy) long wavelength configurations of the field $\phi$,
only this direction can be excited and we write
$$	\phi \simeq v e^{ia/f} \,\, , \eqno (2.13)$$
where $a$ is the Goldstone boson field and $f$ is a constant.  The field
$a$ corresponds to excitations that cost no potential energy. Under a $U(1)$
transformation
$$	a/f \rightarrow a/f + \Omega \,\, . \eqno (2.14)$$
Putting (2.13) into eq. (2.10) gives
$$	{\cal L} = (v/f)^2 \partial_\mu a \partial^\mu a \,\, , \eqno
(2.15)$$
so for a properly normalized kinetic term the constant $f$ is given by
$$	f = \sqrt{2} v \,\, . \eqno (2.16)$$
The symmetry, eq. (2.14), insures that the Lagrange density, eq. (2.15), has no
mass term for $a$.

An analysis very similar to the above holds for QCD.  The Goldstone
bosons are included in a $3 \times 3$ special unitary matrix $\Sigma_{ba}$
that can be thought of as arising from transforming $\bar q_{aR} q_{bL}$
away from its vacuum expectation value.  In analogy with the $U(1)$ case
for long-wavelength (low-energy) excitations we write
$$	\bar q_{aR} q_{bL} \simeq v \Sigma_{ba}\,\, . \eqno (2.17)$$
Under an $SU(3)_L \times SU(3)_R$ transformation the above
identification implies that
$$	\Sigma \rightarrow L\Sigma R^{\dag} \,\, . \eqno (2.18)$$
The matrix $\Sigma$ is the analog of $e^{ia/f}$ in the $U(1)$ case.  To
display explicitly the Goldstone boson fields we write
$$	\Sigma = \exp \left({2iM\over f}\right) \eqno (2.19)$$
where
$$	M = \left[ \matrix{{1\over\sqrt{2}} \pi^{0} + {1\over\sqrt{6}} \eta
& \pi^+ & K^+\cr
\pi^- &  - {1\over\sqrt{2}} \pi^0 + {1\over\sqrt{6}} \eta & K^0\cr
K^- & \bar K^0 & {-2\over\sqrt{6}} \eta\cr} \right] \,\, . \eqno
(2.20)$$
In eq. (2.19) $f$ is a constant with dimensions of mass.  It is
straightforward to see using eq. (2.17) that the particle assignments above are
correct.
Note that under an unbroken $SU(3)_V$ transformation $L = R = V$, eq.
(2.18) implies that
$$	M \rightarrow VMV^{\dag}\,\, , \eqno (2.21)$$
so the $\pi, K$, and $\eta$ transform as an $SU(3)_V$ octet.

\noindent {\bf 3.  An Effective Lagrangian for the Strong Interactions
of the Goldstone Bosons}

To describe the strong interactions of the $\pi, K$ and $\eta$ at low
momentum an effective chiral Lagrangian is constructed.  This effective
Lagrangian contains only these fields (heavier degrees of freedom, e.g.,
the $\rho$-meson, have been integrated out), and because we are
interested in low-momentum physics the Lagrangian can be expanded in
derivatives.  Terms with more derivatives are suppressed since
derivatives bring down factors of the small momentum.  We want the Lagrangian
to
respect the symmetries of QCD, i.e., parity and charge conjugation.  If
for the moment the light quark masses are neglected, the Lagrangian
should also be invariant under chiral $SU(3)_L \times SU(3)_R$.  Thus
$$	{\cal L} = {f^2\over 8} Tr \partial_\mu \Sigma \partial^\mu
\Sigma^{\dag} + ... \,\, \eqno (3.1)$$
where the ellipsis denotes terms with more than two derivatives.  Note
that there are no terms with zero derivatives because $Tr \Sigma^{\dag}
\Sigma = 1$.  The factor $f^2/8$ is inserted to get properly normalized
kinetic terms for the Goldstone bosons.

So far we haven't included the quark mass terms.  Recall
$$	{\cal L}_{mass} = - (\bar q_L m_q q_R + \bar q_R m_q q_L) \,\,
,\eqno (3.2)$$
and under a $SU(3)_L \times SU(3)_R$ transformation
$$	{\cal L}_{mass} \rightarrow - (\bar q_L L^{\dag} m_q R q_R + \bar q_R
R^{\dag} m_q L q_L)\,\, . \eqno (3.3)$$
Hence the quark mass terms transform as $(\bar 3_L, 3_R) + (3_L, \bar
3_R)$.  If we add to eq. (3.1) the most general terms that transform this
way then we have included, to first order in $m_q$, the effects of the
quark masses.  There is a very simple way to do the group theory.  If we
pretend that the quark mass matrix transforms as $m_q \rightarrow Lm_q
R^{\dag}$ and construct invariants under $SU(3)_L \times SU(3)_R$ then
the effects of the $(\bar 3_L, 3_R)$ term in eq. (3.2) are taken into
account.  Similarly if we pretend that the quark mass matrix transforms as $m_q
\rightarrow Rm_q L^{\dag}$ and construct invariants under $SU(3)_L
\times SU(3)_R$ the effects of the $(3_L, \bar 3_R)$ term in eq. (3.2)
are taken into account.  Including terms linear in the light quark mass
matrix, the chiral Lagrange density becomes
$$	{\cal L} = {f^2\over 8} Tr \partial_\mu \Sigma \partial^\mu
\Sigma^{\dag} + \lambda_o Tr (m_q \Sigma + \Sigma^{\dag} m_q)\,\, ,\eqno
(3.4)$$
where $\lambda_o$ is a constant with dimension $(mass)^3$.  The
quark mass terms give masses to the Goldstone bosons (i.e., they are
only approximate Goldstone bosons because $SU(3)_L \times SU(3)_R$
symmetry is not exact)
$$	\eqalignno{m_{\pi^{\pm}}^2 &= {4\lambda_o\over f^2} (m_u + m_d)
	&(3.5a)\cr
	m_{K^{0}}^2 &= {4\lambda_o\over f^2} (m_d + m_s)& (3.5b)\cr
	m_{K^{\pm}}^2 &= {4\lambda_o\over f^2} (m_u + m_s)\,\, ,& (3.5c)\cr}$$
and for the $\pi^0$ and $\eta$ there is a $2 \times 2$ mass matrix
$$	m_{(\pi^{0},\eta)}^2 = {4\lambda_o\over f^2} \left[\matrix{(m_u
+ m_d) & {1\over \sqrt{3}} (m_u - m_d)\cr
{1\over \sqrt{3}} (m_u - m_d) & {1\over 3} (m_u + m_d) + {4\over 3}
m_s\cr}\right]\,\, .\eqno (3.6)$$

Since $m_K^2/m_\pi^2 \simeq m_s/(m_u + m_d)$ is large, the strange quark
mass is much larger than the up and down quark masses.
The eigenvalues of the matrix in eq. (3.6) are then given approximately by the
diagonal elements;  corrections to this from the off diagonal elements
are of order $((m_u - m_d)/m_s)^2$.  Neglecting $m_{u,d}$ compared with
$m_s$ gives the relation
$$	(3/4)m_\eta^2 = m_K^2 \,\, . \eqno (3.7)$$
An examination of isospin splittings leads to the expectation that $m_d$
is about twice as large as $m_u$.  Despite the fact that $m_d/m_u$
differs significantly from unity, the neutral and charged pions are
almost degenerate because the strange quark mass is much larger than
both the up and down quark masses.

The transformation properties of $\Sigma$ under charge conjugation
$$	C\Sigma C^{-1} = \Sigma^T \,\, ,\eqno (3.8)$$
and parity
$$	P \Sigma (\vec x,t) P^{-1} = \Sigma^{\dagger} (- \vec x, t) \,\,
, \eqno (3.9)$$
follow from eq. (2.17).

Under an infinitesimal space-time-dependent left-handed transformation
$$	L \simeq 1 + i \epsilon_L^A T^A \,\, , \eqno (3.10)$$
the QCD Lagrange density (2.6) changes by
$$	\delta {\cal L} = - L_\mu^A \partial^\mu \epsilon_L^A \,\, ,
\eqno (3.11)$$
where
$$	L_\mu^A = \bar q_L T^A \gamma_\mu q_L \,\, , \eqno (3.12)$$
is the left-handed current.  On the other hand, under the transformation
eq. (3.10), the chiral Lagrange density, eq. (3.1), changes by
$$	\delta {\cal L} =  {-if^2\over 4} Tr (\partial_\mu \Sigma
\Sigma^{\dagger} T^A) \partial^\mu \epsilon_L^A \,\, . \eqno (3.13)$$
Comparison of eqs. (3.11) and (3.13) yields an expression for the left-handed
current in terms of $\pi, K$ and $\eta$ fields
$$	L_\mu^A = {if^2\over 4} Tr (\partial_\mu \Sigma \Sigma^{\dagger}
T^A) \,\, . \eqno (3.14)$$

At higher orders in derivatives this Noether procedure for extracting
the symmetry currents become ambiguous.  Terms in the Lagrange density
that are total derivatives contribute to the currents (but not the
charges.)  An alternate method for deriving the symmetry currents is to
gauge the chiral symmetries.  Then, for example, the coupling of the
external left-handed gauge field, $A_L^\mu$ is proportional to $A_L^{C\mu}
L_\mu^C$.  In this procedure the ambiguity associated with total derivatives
corresponds to terms in the gauged chiral Lagrangian that contain a
field strength tensor.

Low-momentum matrix elements of the left handed current, eq. (3.12),
involving $\pi, K$ and $\eta$ fields are given by the matrix elements of
eq. (3.14).  Expanding $\Sigma$ in terms of $M$ gives
$$	L_\mu^A = {-f\over 2} Tr \partial_\mu MT^A + ... \,\, ,\eqno (3.15)$$
where the ellipsis denote higher powers of $M$.

The invariant matrix element for $\pi^-(p) \rightarrow \mu (p_\mu) \bar
\nu_\mu (p_\nu)$ decay is
$$	{\cal M} = {G_F\over\sqrt{2}} c_1 <0|\bar u \gamma_\mu (1 - \gamma_5)
d| \pi^- (p_\pi)>$$
$$	\cdot \bar u(p_\mu) \gamma^\mu (1 - \gamma_5) v(p_\nu) \,\, ,
\eqno (3.16)$$
where $c_1$ is the cosine of the Cabibbo angle.
But according to eqs. (3.12) and (3.15), $\bar u \gamma_\mu (1 - \gamma_5) d
= - f \partial_\mu \pi^- + ...$, so eq. (3.16) becomes
$$	\eqalign{{\cal M} &= i {G_F c_1 f\over \sqrt{2}} p^\mu_\pi \bar u
(p_\mu) \gamma_\mu (1 - \gamma_5) v(p_\nu)\cr
& = i {G_F c_1 f\over \sqrt{2}} m_\mu \bar u (p_\mu)(1 - \gamma_5)
v(p_\nu)\,\, .\cr} \eqno (3.17)$$
Comparing the rate that results from eq. (3.17) with experiment gives $f
\simeq$ 132 MeV.

\noindent {\bf 4.  Power Counting for $\pi \pi$ Scattering}

Chiral perturbation theory can be used to predict the cross section for
$\pi \pi$ scattering at low momentum.  Taking Lagrange density
(3.4) and expanding $\Sigma$ in powers of $M$ the pieces with four $M$'s
give the tree level contribution to the invariant matrix element for the
$\pi \pi \rightarrow \pi \pi$ process.  The lowest four-momentum $p$
accessible is of order $m_\pi$.  Since $m_\pi^2$ is linear in the light
quark masses, an insertion of $m_q$ is of the same order as two
derivatives.  At tree level the Lagrange density, eq. (3.4), gives an
invariant matrix element of order $p^2/f^2$.

One-loop Feynman diagrams like that in Figure 1 also contribute to the
invariant matrix element for $\pi \pi$ scattering.  To describe their
contribution it is convenient to pick dimensional regularization with
minimal subtraction as the renormalization scheme.  This is particularly
convenient because there is no dimensionful cutoff and the subtraction
point $\mu$ only appears in the argument of logarithms (e.g., as $\log
(p^2/\mu^2)$).  The Feynman diagram in Figure 1 contains four factors of
$1/f$ (two from each vertex) and is of order $(1/16 \pi^2) (p^4/f^4)
\ln (p^2/\mu^2)$.  Hence, it is subdominant to the tree level
contribution of the two derivative term in the chiral Lagrangian, and
roughly comparable in importance with the tree level
contribution of terms in the chiral Lagrangian containing four
derivatives. Such higher-dimension terms in the Lagrangian have coefficients
that depend on $\mu$, and this dependence cancels the logarithmic
dependence on $\mu$ from the one-loop diagrams$^{[5]}$.
For small $p^2$, and $\mu$ of order the chiral
symmetry breaking scale $\sim 1 GeV$, the one-loop contribution is enhanced
by a large logarithm
over the tree level contribution of the terms in the Lagrangian with
four derivatives.

Higher loops will contain more powers of $(1/f)$ and
are even less important at low momentum.  So we have seen that at low
momentum the tree level contribution of the terms in the Lagrange density
with two derivatives or one factor of the light quark mass matrix
dominate the $\pi\pi \rightarrow \pi\pi$ matrix element.  This matrix
element is predicted just in terms of $f$ and the pion mass.

Similar power counting applies to other processes.  For example, for the
matrix element $<0|L_\nu^A|\pi>$ the contributions from loops and
higher-derivative terms in the chiral Lagrangian are suppressed compared
to the
contribution evaluated in the previous section.

\noindent {\bf 5.  Semileptonic Kaon Decay}

In this section chiral perturbation theory is applied to the
semileptonic decays$^{[6]}$ $K^0 \rightarrow \pi^- e^+ \nu_e$ and $K^0
\rightarrow
\pi^0 \pi^- e^+ \nu_e$.  At the quark level the effective Hamiltonian
density for these decays is
$$	H_{eff} = {G_F \over\sqrt{2}} s_1 (\bar s \gamma_\mu (1 -
\gamma_5) u) (\bar\nu_e \gamma^\mu (1 - \gamma_5) e) \,\, . \eqno
(5.1)$$

The amplitude for $K^0 \rightarrow \pi^- e^+ \nu_e$ decay is dependent
on the hadronic matrix element
$$	<\pi^- (p_\pi) |\bar s \gamma_\mu (1 - \gamma_5) u|K^0 (p_K)> =
f_+ (p_K + p_\pi)_\mu + f_- (p_K - p_\pi)_\mu \,\, , \eqno (5.2)$$
where $f_\pm$ are (Lorentz invariant) functions of the square of the
momentum transfer $q^2 = (p_K - p_\pi)^2$.  According to eqs. (3.12)
and (3.14), for low energy matrix elements involving the $\pi, K$ and
$\eta$ we can write
$$	\bar s \gamma_\mu (1 - \gamma_5) u = {if^2\over 2} Tr
(\partial_\mu \Sigma \Sigma^{\dagger} T) \,\, , \eqno (5.3)$$
with
$$	T = \pmatrix{0 & 0 & 0\cr
	0 & 0 & 0\cr
	1 & 0 & 0\cr} \,\, . \eqno (5.4)$$
Expanding $\Sigma$ in terms of $M$ in eq. (5.3), terms with two $M$'s
lead to the prediction
$$	f_+ = 1 ~~, \qquad f_- = 0 \,\, . \eqno (5.5)$$

For $K^0 \rightarrow \pi^0 \pi^- e^+ \nu_e$ decay the relevant hadronic
matrix element is
$$	<\pi^- (q) \pi^0 (k) |\bar s \gamma^\mu (1 - \gamma_5) u|K^0 (p)>$$
$$	= iF_1 (q + k)^\mu + iF_2 (q - k)^\mu + iF_3 (p - q - k)^\mu + F_4
\epsilon^{\mu\nu\lambda\sigma} p_\nu q_\lambda k_\sigma \,\, . \eqno
(5.6)$$
The terms with three $M$'s in eq. (5.3) contribute (Figure 2a) to this
matrix element as well
as the pole graph in Figure 2b.  (Isospin-violating effects are neglected
here.)   In Figure 2b shaded circle represents a strong
interaction vertex while a shaded square denotes an insertion of the
left-handed current.  The two derivatives in the strong vertex produce two
factors of momentum which are cancelled (so far as the power counting is
concerned) by the propagator.  Consequently both diagrams in Figure 2
contribute to the matrix element (5.6) in the leading order of chiral
perturbation theory and give
$$	F_1 = 0,\quad F_2 = - {\sqrt{2}\over f},\quad F_3 = {\sqrt{2}\over f} {(q
- k) \cdot (p - q - k)\over (m_K^2 - (p - q - k)^2)}\,\, .
\eqno (5.7)$$
The leading contribution to the form factor $F_4$ arises from the Wess--Zumino
term.  This is discussed in the next section.

\noindent {\bf 6.  The Wess--Zumino Term}

The currents $L_\mu^A$ and $R_\mu^A$ associated with chiral $SU(3)_L
\times SU(3)_R$ symmetry are exactly conserved in the $m_q = 0$ limit.
However, quark loops give an anomalous contribution to matrix
elements involving these currents (e.g., the contribution to
${\partial\over\partial x^\mu} <0|T(L^{A\mu} (x) L^{B\nu} (y) L^{C\alpha}
(z))|0>$ from the usual triangle diagram).  The chiral
Lagrangian of Section 3 does not have these anomalous contributions.  A
simple way to include the anomalous effects of quark loops on the $\pi,
K$ and $\eta$ interactions is to add to the chiral Lagrange density
massive left-and right-handed fermion fields $\hat q_{aL}$ and $\hat
q_{aR}$ that transform under chiral symmetry the same way the quarks do:
$\hat q_L \rightarrow L \hat q_L, \hat q_R \rightarrow R \hat q_R$.  The
Lagrange density for these fields is
$$	{\cal L} = i {~\bar{\!\hat q}}_L \parslash \hat q_L + i
{~\bar{\!\hat q}}_R \parslash \hat q_R - M_q({~\bar{\!\hat q}}_L
\Sigma \hat q_R + {~\bar{\!\hat q}}_R \Sigma^{\dagger} q_L)\,\, . \eqno (6.1)
$$
Integrating out these massive fermions will reproduce anomalous effects
of quark loops on $\pi, K$, and $\eta$ interactions.  Note that the $\pi,
K$ and $\eta$ have $SU(3)_L \times SU(3)_R$ invariant nonderivative couplings
to $\hat q_{L,R}$
proportional to $M_q$.  Large $M_q$ factors from vertices can
cancel those from propagators.  It is these nonderivative couplings
proportional to $M_q$ that are responsible for the anomalous  $\pi, K$ and
$\eta$ interactions.

Now suppose we make a field redefinition
$$	\eqalignno{\hat q_L(s) &= \Sigma^{\dagger} (s) \hat q_L & (6.2a)\cr
\hat q_R (s) &= \hat q_R \,\, , & (6.2b)\cr}$$
where $\Sigma(s)$ is an element of a continuous one-parameter family of special
unitary matrices that satisfies
$$	\Sigma(0) = 1 \qquad \qquad \Sigma (1) = \Sigma \,\, .\eqno (6.3)$$
In terms of $\hat q_{L,R}(s)$ the action becomes
$$	\eqalign{S &= \int d^4 x [{\bar{\hat q}}_L (s) (i \parslash
- \Aslash_L) \hat q_L(s) + {\bar{\hat q}}_R (s) i \parslash {\bar{\hat
q}}_R(s)\cr
&- M_q({\bar{\hat q}}_L(s) \Sigma^{\dagger} (s) \Sigma \hat q_R(s) +
{\bar{\hat q}}_R (s) \Sigma^{\dagger} \Sigma (s) \hat q_L (s))]\cr
&+ \Delta (s)\,\, ,\cr} \eqno (6.4)$$
where $\Delta (s)$ comes from the anomaly (i.e., from the Jacobian associated
with the transformation in eq. (6.2)) and $A_L^\mu$ is
$$	A_L^\mu = i \Sigma^{\dagger} (s) \partial^\mu \Sigma (s) \,\, .
\eqno (6.5)$$

At $s = 1$ the full effect of the anomaly on $\pi, K$ and $\eta$
interactions is in $\Delta (1)$ since there are no longer nonderivative
couplings of the Goldstone bosons to the heavy fermions that are proportional
to $M_q$.  It is straightforward using the Noether procedure to compute how
$\Delta$ changes with $s$.  Under a change from $s$ to $s + \delta s$
$$	\delta \hat q_L (s) = i \epsilon \hat q_L (s) \,\, , \quad
\delta \hat q_R (s) = 0\,\, , \eqno (6.6)$$
where
$$	\epsilon (s) = - i (\delta s) \partial_s \Sigma^{\dagger} (s)
\Sigma \,\, . \eqno (6.7)$$
The change in $\Delta (s)$ from the anomaly is
$$	\delta \Delta (s) = \int d^4 x Tr \epsilon (D_\mu
L^\mu)_{{\rm anomalous}}\,\, , \eqno (6.8)$$
where the anomalous divergence of the current\foot{The current $L^\mu$
is the matrix with components $L_{ab}^\mu = \bar q_{aL} \gamma^\mu
q_{bL}$.} $L^\mu$ is
$$	(D^\mu L_\mu)_{{\rm anomalous}} = {-1\over 24\pi^2}
\epsilon^{\mu\nu\rho\sigma}
[\partial_\mu A_\nu^L \partial_\rho A_\sigma^L -
	{i\over 2} \partial_\mu (A_\nu^L A_\rho^L A_\sigma^L)] \,\, ,
\eqno (6.9)$$
since $A_\nu^L$ appears as an external ``gauge field'' in eq. (6.4).  With
$A_\nu^L$
given by eq. (6.5), eq. (6.9) becomes
$$	(D^\mu L_\mu)_{\rm anomalous} = {1\over 48\pi^2} \epsilon^{\mu\nu\rho\sigma}
\partial_\mu \Sigma^{\dagger} \partial_\nu \Sigma \partial_\rho
\Sigma^{\dagger} \partial_\sigma \Sigma \,\, . \eqno (6.10)$$
So as $s$ changes from $s$ to $s + \delta s$,
$$	\delta \Delta (s) = \delta s {i\over 48\pi^2} \int d^4 x
\epsilon^{\mu\nu\rho\sigma}
	Tr (\Sigma^{\dagger} \partial_s \Sigma \partial_\mu
\Sigma^{\dagger} \partial_\nu \Sigma \partial_\rho\Sigma^{\dagger}
\partial_\sigma \Sigma) \,\, . \eqno (6.11)$$
Integrating this from $s = 0$ to $s = 1$ gives the effect of the anomaly
on $\pi, K$ and $\eta$ interactions.  This is the Wess--Zumino
term$^{[7,8]}$
$$	S_{WZ} = 3 \Delta (1)$$
$$	= {i\over 16\pi^2} \int\nolimits^1_0 ds \int d^4 x
\epsilon^{\mu\nu\rho\sigma} Tr [\Sigma^{\dagger} \partial_s \Sigma
\partial_\mu \Sigma^{\dagger} \partial_\nu \Sigma \partial_\rho
\Sigma^{\dagger} \partial_\sigma \Sigma] \,\, . \eqno (6.12)$$
The factor of three is inserted in eq. (6.12) because $\Delta (1)$ only
includes the effects of a single color of up, down and strange quarks.

The integration region in (6.12) is over a five-dimensional disc of
which spacetime is the boundary (see eq. 6.3) at $s = 1$.  Using $x^5 =
s$ the Wess--Zumino term can be written in the more symmetrical form
$$	S_{WZ} = {i\over 80\pi^2} \int_D d^5 x
\epsilon^{\mu\nu\rho\sigma\alpha} Tr [\Sigma^{\dagger} \partial_\mu
\Sigma \partial_\nu \Sigma^{\dagger} \partial_\rho \Sigma \partial_\sigma
\Sigma^{\dagger} \partial_\alpha \Sigma]\,\, . \eqno(6.13)$$
When $\Sigma$ is expanded in terms of $M$ the various terms can be
written as integrals over four-dimensional spacetime.  However, the
$SU(3)_L \times SU(3)_R$ invariance of the Wess--Zumino term is only
manifest when it is written as an
integral over a five-dimensional disk.

The Wess--Zumino term contributes to the left-handed current
$$	L^{\mu A}_{WZ} = {1\over 16\pi^2} \epsilon^{\mu\nu\rho\sigma} Tr
\partial_\nu \Sigma \partial_\rho \Sigma^{\dagger} \partial_\sigma \Sigma
\Sigma^{\dagger} T^A \,\, . \eqno (6.14)$$
It is this anomalous piece of the left-handed current that is
responsible for the leading contribution to the form factor $F_4$ in
$K^0 \rightarrow \pi^- \pi^0 e^+ \nu_e$ decay.  This is because
other terms in the strong interaction Lagrange density with four
derivatives and an antisymmetric tensor (e.g., $Tr
\epsilon^{\mu\nu\lambda\sigma} \Sigma^{\dagger} \partial_\mu \Sigma
\Sigma^{\dagger} \partial_\nu \Sigma \Sigma^{\dagger} \partial_\lambda
\Sigma \Sigma^{\dagger} \partial_\sigma \Sigma$) are forbidden by
parity.  Expanding eq. (6.14) out in powers of $M$, the piece with three
$M$'s gives
$$	F_4 = {\sqrt{2}\over \pi^2 f^3}\,\, . \eqno (6.15)$$

\noindent {\bf 7.  Heavy Quark Symmetries of the Strong Interactions}

The part of the Lagrange density for QCD involving the heavy quark field $Q$
is
$$	{\cal L} = \bar Q (i\Dslash - m_Q)Q \,\, . \eqno (7.1)$$
We shall be interested in the kinematic situation where the heavy quark
is interacting with light degrees of freedom (i.e., gluons and light
quarks and antiquarks)  carrying momenta that are typically much
smaller than the heavy quark mass.  In this situation it is appropriate
to take the limit of QCD where the heavy quark mass $m_Q$ goes to
infinity with its four-velocity $v^\mu$ fixed.  To do this we write for
the heavy quark with velocity $v$
$$	Q = e^{-i m_Q v \cdot x} [h_v^{(Q)} + \chi_v^{(Q)}] \,\, , \eqno (7.2)$$
where
$$	\vslash h_v^{(Q)} = h_v^{(Q)}, \quad \vslash \chi_v^{(Q)} = -
\chi_v^{(Q)} \,\, . \eqno (7.3)$$

In the kinematic situation of interest the heavy quark is almost
on shell and so $\chi_v^{(Q)}$ can be treated as a small quantity.
$h_v^{(Q)}$ and $\chi_v^{(Q)}$ have a much less rapid dependence on
spacetime than the phase factor explicitly factored out in eq. (7.2).
Neglecting $\chi_v^{(Q)}$ and substituting eq. (7.2) into (7.1) we find
$$	{\cal L} = \bar h_v^{(Q)} [m_Q (\vslash - 1) + i\Dslash] h_v^{(Q)}$$
$$	= \bar h_v^{(Q)} i \Dslash h_v^{(Q)} \,\, . \eqno (7.4)$$
This can be further simplified using eq. (7.3) to
$$	{\cal L}_v = \bar h_v^{(Q)} \left({\vslash + 1 \over 2}\right) i
\Dslash h_v^{(Q)}$$
$$	= \bar h_v^{(Q)} \left[ i v \cdot D - i \Dslash \left({\vslash -
1\over 2}\right)\right] h_v^{(Q)}$$
$$	= \bar h_v^{(Q)} i v \cdot D h_v^{(Q)} \,\, . \eqno (7.5)$$
The Lagrange density for the effective heavy quark theory in eq. (7.5) has as
its Feynman rules: $i/(v \cdot k + i\epsilon$) for the heavy quark
propagator,
and $-ig T^A v_\mu$ as the vertex for the gluon heavy quark interaction.
The full four-momentum of  the heavy quark is $p_Q = m_Q v + k$.  The
momentum $k$ that occurs in the heavy quark propagator of the effective
theory is called the residual momentum;  it is a measure of how much the
heavy quark is off-shell.  For the heavy quark effective theory to be
valid $k$ must be much less than $m_Q$.

Note that in the effective theory the field $h_v^{(Q)}$ destroys a heavy
quark of four-velocity $v$; it does not create the corresponding
antiquark.  Pair creation of heavy quarks does not occur in the
effective theory.  The equation of motion for the field $h_v^{(Q)}$ is
$$	v \cdot Dh_v^{(Q)} = 0 \,\, . \eqno (7.6)$$

The heavy quark effective theory has symmetries that are not manifest in
the full theory of QCD.$^{[9, 10]}$
Since there is no pair creation in the effective theory, there is a $U(1)$
symmetry of the Lagrange density in eq. (7.5) associated with heavy quark
conservation.    Under an infinitesimal $U(1)$ transformation of this
type
$$	h_v^{(Q)} \rightarrow h_v^{(Q)} + \delta h_v^{(Q)} \,\, , \eqno
(7.7)$$
with
$$	\delta h_v^{(Q)} = i \epsilon_0 h_v^{(Q)}\,\, . \eqno (7.8)$$
Here $\epsilon_0$ is an arbitrary (real) infinitesimal parameter.
Since gamma matrices no longer occur in the effective theory,
the spin of the heavy quark is conserved by the heavy quark-gluon
interaction.  Associated with this is an
$SU(2)$ symmetry group of the Lagrange density in eq. (7.5).  To define the
action of the $SU(2)$ group on the heavy quark fields, we introduce three
orthonormal four-vectors $e_{a\mu}, a = 1,2,3$, that are orthogonal to the
heavy quark's four-velocity:
$$	e_{a\mu} e_b^\mu = - \delta_{ab}\,\, , \eqno (7.9)$$
$$	v_\mu e_a^\mu = 0  \,\, . \eqno (7.10)$$
Then the three $4 \times 4$ matrices
$$	S^a = {i\over 8} \sum_{b,c} \epsilon^{abc} [\eeslash_b,
\eeslash_c] \,\, , \eqno (7.11)$$
are the generators of the heavy quark spin symmetry.  The Lagrange
density in eq. (7.5) is invariant under the infinitesimal transformations
$$	h_v^{(Q)} \rightarrow h_v^{(Q)} + \delta h_v^{(Q)} \,\, , \eqno
(7.12)$$
where
$$	\delta h_v^{(Q)} = i \sum_a \epsilon^a S^a h_v^{(Q)}\,\, . \eqno
(7.13)$$
Furthermore, $[\vslash, S^a] = 0$.  So these transformations preserve
the constraint $\vslash h_v^{(Q)} = h_v^{(Q)}$.  It is easy to see that these
transformations correspond to heavy quark spin symmetry
in the heavy quark rest frame  where $v^\mu = (1,\vec 0)$.  In
this frame $h_v^{(Q)}$ is a two-component Pauli spinor (the lower two
components vanish because of the constraint $\gamma^0 h_v^{(Q)} =
h_v^{(Q)}$).  Using the explicit representation
$$	\gamma^j = \pmatrix{0 & \sigma^j\cr
	-\sigma^j & 0 \cr}\,\, , \eqno (7.14)$$
and picking the $e_{a\mu}$ to be unit vectors along the three spatial
axes we find that eq. (7.11) implies $S^a = {1\over 2} \pmatrix{\sigma^a & 0\cr
0 & \sigma^a\cr}$.

If there are $N_h$ heavy quarks $Q_{1} ... Q_{N_{h}}$ moving with the same
four-velocity $v$, then, denoting the corresponding fields in the
effective theory by $h_v^{(i)}$, the Lagrange density becomes
$$	{\cal L}_v = \sum_{j = 1}^{N_{h}} \bar h_v^{(j)} i v \cdot
Dh_v^{(j)} \,\, . \eqno (7.15)$$

The Lagrange density in eq. (7.15) is independent of the heavy quark
masses, and so the $SU(2)$ spin symmetry generalizes to a $SU(2N_h)$
spin-flavor symmetry.  Note that because the heavy quark masses can be
very different, this symmetry  relates quarks of the same four-velocity
but generally  different momentum.  This is one of the
unusual aspects of heavy quark symmetry.

In nature there are three heavy quarks, $c,b$ and $t$, and so the heavy
quark spin-flavor symmetry is $SU(6)$.  However, because the top quark
is so heavy it is likely to decay before it forms a hadron.  Ironically,
heavy quark symmetry is not a very useful concept for the heaviest of all
quarks.

Heavy quark symmetry is not just a nonrelativistic symmetry.  In our
derivation of the effective theory the spatial components of the
four-velocity are not necessarily small.  With just a single heavy
quark interacting strongly this is not a particularly significant
statement, since one can always choose to work in the rest frame where
$v^\mu = (1,\vec 0)$.  However, we will be interested in cases where an
external source (e.g., a $W$-boson) takes a heavy quark and changes it
into a heavy quark with a different four-velocity (and possibly a
different flavor as well).  Then it is not possible to be in the
rest frame of both the initial and final quarks.

\noindent {\bf 8.  Heavy Hadron Multiplets}

In the $m_Q \rightarrow \infty$ limit the total angular momentum of the
light degrees of freedom
$$	\vec S_\ell = \vec S - \vec S_Q \,\, , \eqno (8.1)$$
commutes with the Hamiltonian.  Thus $s_\ell$, the angular momentum of
the light degrees of freedom in the hadron's rest frame, is a good quantum
number.$^{[11]}$  Consequently in the $m_Q \rightarrow \infty$ limit hadrons
containing a single heavy quark come in degenerate doublets of total
spin
$$	s_\pm = s_\ell \pm 1/2 \,\, , \eqno (8.2)$$
unless $s_\ell = 0$, in which case the total spin is $s = 1/2$.  For example
when $s_\ell = 1/2$, there are spin-zero and spin-one states
$$	\eqalignno{|0> &= {1\over \sqrt{2}} [|\uparrow \downarrow > - |
\downarrow \uparrow>] & (8.3a)\cr
	|1, 1> &= | \uparrow \uparrow> & (8.3b)\cr
	|1, 0> &= {1\over \sqrt{2}} [| \uparrow \downarrow > + |
\downarrow \uparrow >] & (8.3c)\cr
	|1, -1> &= | \downarrow \downarrow > & (8.3d)\cr}$$
where the first arrow represents the heavy quark spin (along the 3rd axis)
while the second arrow denotes the spin of the light degrees of freedom.  Since
$$	S_Q^3   |0> = {1\over 2} |1,0>\,\, , \eqno (8.4)$$
and $\vec S_Q$ commutes with the Hamiltonian, these spin-zero and
spin-one states are degenerate in mass.

For mesons with $Q \bar q_a$ flavor quantum numbers, the ground state
multiplet has $s_\ell = 1/2$ and negative parity, corresponding to the
pseudoscalar mesons $P_a$ and vector mesons $P_a^*$.  For $Q = c$
these are the $(D^0, D^+, D_s)$ and $(D^{*0}, D^{*+}, D_s^*)$
mesons while for $Q = b$ these are the $(B^-, B^0, B_s)$ and $(B^{*-},
B^{*0}, B_s^*)$ mesons.  We denote the fields that destroy hadrons of
this type, with four-velocity $v$, by $P_a$ and $P_a^{*\mu}$.  The
vector field satisfies the constraint
$$	v_\mu P_a^{*\mu} = 0 \,\, . \eqno (8.5)$$
It is convenient to combine these fields into a $4 \times 4$ matrix
$H_a$ in the following fashion$^{[3]}$
$$	H_a = \left({\vslash + 1\over 2}\right) [P_a^{*\mu} \gamma_\mu -
P_a \gamma_5] \,\, . \eqno (8.6)$$
This is a shorthand notation.  In cases where the flavor of the heavy
quark $Q$ and the value of the four-velocity $v^\mu$ are important we
shall use $H_a^{(Q)} (v)$.  One can think of the relationship between
$H_a$ and the underlying degrees of freedom schematically as
$$	h_v^{(Q)} \bar \ell_a \sim H_a^{(Q)} (v) \,\, , \eqno (8.7)$$
where $\bar\ell_a$ is a spinor field that destroys the light degrees of
freedom.  Eq. (8.7) implies that with respect to Lorentz transformations
$H_a(v) \rightarrow D(\Lambda^{-1}) H_a (\Lambda v) D
(\Lambda^{-1})^{-1}$, where $D(\Lambda)$ is
the usual $4 \times 4$ Dirac representation of the Lorentz group (i.e.,
$H_a (v)$ transforms as a bispinor).  This transformation law gives
$P^{*\nu} \rightarrow \Lambda^\nu_{~\mu} P^{*\mu}$ because $D(\Lambda^{-1})
\gamma^\nu D(\Lambda^{-1})^{-1} = \Lambda^\nu_{~\mu} \gamma^\mu$.  $H_a$
transforms under heavy quark spin symmetry $SU_(2)_v$ as
$$	H_a \rightarrow SH_a \,\, ,\eqno (8.8)$$
where $S$ is an element of $SU(2)_v$.  In the heavy meson rest frame $v^\mu
= (1,\vec 0)$ eq. (8.8) becomes
$$	\delta P_a = {1\over 2} i \epsilon^k P_a^{*k}\,\, , \eqno (8.9)$$
$$	\delta P_a^{*k} = {1\over 2} i\epsilon^k P_a - {1\over 2}
\epsilon^{j\ell k} \epsilon^j P_a^{*\ell}\,\, , \eqno (8.10)$$
for infinitesimal transformations $S = 1 + i\sum \epsilon^j S_Q^j$.  Eq.
(8.9) corresponds to eq. (8.4) and the analogous equations that result from
applying the heavy quark spin raising and lowering operators $S_Q^\pm$
to the spin-zero state.

The matrix $H_a^{(Q)} (v)$ satisfies the identities
$$	\vslash H_a^{(Q)} (v) = H_a^{(Q)} (v) \eqno (8.11a)$$
$$	H_a^{(Q)} (v) \vslash = - H_a^{(Q)} (v)\,\,. \eqno (8.11b)$$
It is convenient to introduce $\bar H_a = \gamma^0 H_a^{\dagger} \gamma^0$.
Under Lorentz transformations $\bar H_a \rightarrow D(\Lambda^{-1}) \bar H_a
D(\Lambda^{-1})^{-1}$, while under heavy quark spin symmetry $\bar H_a
\rightarrow
\bar H_a S^{-1}$.

Some excited mesons with $Q \bar q_a$ flavor quantum numbers have been
observed (for the case $Q = c$).  In the nonrelativistic constituent
quark model, the lowest mass excitations arise from giving the light
antiquark a unit of orbital angular momentum.  This results in two
positive parity multiplets of heavy mesons, one with $s_\ell = 1/2$ and
the other with $s_\ell = 3/2$.

For $s_\ell = 1/2$ we denote the fields that destroy the spin-zero and
spin-one positive parity mesons in this multiplet by $P_a^*$ and $P_a^{'\mu}$
respectively  $(v_\mu
P_a^{'\mu} = 0$).  Again it is convenient to combine them into a $4
\times 4$ matrix
$$	G_a = {(1 + \vslash)\over 2} (P_a^{'\mu} \gamma_\mu \gamma_5 - P_a^*)
\,\, , \eqno (8.12)$$
that transforms under heavy quark spin symmetry as
$$	G_a \rightarrow S G_a \,\, , \eqno (8.13)$$
where $S \epsilon SU(2)_v$.  $G_a$ transforms  under Lorentz
transformations in the same way as $H_a$.

For $s_\ell = 3/2$ we denote the fields that destroy the spin-one and
spin-two mesons in this multiplet by $P_a^\mu$ and $P_a^{*\mu\nu}$
respectively  $(v_\mu P_a^\mu = 0, v_\mu P_a^{*\mu\nu} = 0,
P_{a\mu}^{*\mu} = 0$ and $P_a^{*\mu\nu} = P_a^{*\nu\mu})$.  It is convenient
to combine these fields into the $4 \times 4$ matrix$^{[12]}$
$$	F_a^\mu = {(1 + \vslash)\over 2} \left\{P_a^{*\mu\nu} \gamma_\nu -
\sqrt{{3\over 2}}P_a^\nu \gamma_5 \left[g_\nu^\mu - {1\over 3} \gamma_\nu
(\gamma^\mu - v^\mu)\right]\right\} \,\, , \eqno (8.14)$$
that transforms under heavy quark spin symmetry as
$$	F_a^\mu \rightarrow S F_a^\mu \,\, . \eqno (8.15)$$
One can think of the relationship between the
underlying degrees of freedom and $F_a^\mu$ as
$$	h_v^{(Q)} \bar\ell_a^\mu \sim F_a^\mu \,\, , \eqno (8.16)$$
where $\bar\ell_a^\mu$ is a Rarita--Schwinger field that destroys
the spin-3/2 light degrees of freedom.  Eq. (8.16) implies that under
Lorentz transformations $F_a^\mu \rightarrow \Lambda_{~~\nu}^\mu D
(\Lambda^{-1})
F_a^\nu D (\Lambda^{-1})^{-1}$ (the four-velocity also transforms as $v^\mu
\rightarrow \Lambda_{~~\nu}^\mu v^\nu$).

The matrices $G_a$ and $F_a^\mu$ satisfy the relations
$$	\vslash G_a = G_a \qquad \vslash F_a^\mu = F_a^\mu \eqno (8.17a)$$
$$	G_a \vslash = - G_a \qquad F_a^\mu \vslash = - F_a^\mu \eqno
(8.17b)$$
$$	F_a^\mu v_\mu = 0 \qquad F_a^\mu \gamma_\mu = 0 \,\, . \eqno
(8.17c)$$
It is convenient to introduce $\bar G_a = \gamma^0 G_a^{\dagger} \gamma^0$ and
$\bar F_a^\mu = \gamma^0 F_a^{\dagger} \gamma^0$.  These barred fields
transform under Lorentz transformations in the same way as the unbarred
fields.
Under heavy quark spin symmetry $\bar G_a \rightarrow \bar G_a S^{-1}$
and $\bar F_a^\mu \rightarrow \bar F_a^\mu S^{-1}$.

The members of the $s_\ell = 3/2$ multiplet have been observed for $Q
= c$ (and $a = 1$).  They are the $D_1(2420)^0$ and $D_2^* (2460)^0$.
The $Q = c ~(a = 1$ or $2$) members of the $s_\ell = 1/2^+$ multiplet are
expected to have a mass of about 2360 MeV and to be very broad
resonances.  They have not yet been
detected experimentally.

The $4 \times 4$ matrix of fields $H_a$ is an antitriplet with respect
to the unbroken $SU(3)_V$ group.  We need to assign $H_a$ a transformation
rule with respect to the full chiral symmetry group $SU(3)_L \times
SU(3)_R$.  Here, there is considerable freedom associated with our
ability to make field redefinitions.  Suppose $H_a$ transforms as $(\bar 3_L,
1_R)$ under $SU(3)_L \times SU(3)_R$.  Then under an $SU(3)_L \times
SU(3)_R$ transformation
$$	H_a \rightarrow H_b L_{ba}^{\dagger} \,\, , \eqno (8.18)$$
where $L\epsilon SU(3)_L$.  While this is an acceptable transformation
law, it leads to a definition of parity that is somewhat
awkward.  Since parity interchanges left-and right-handed quark fields,
the parity image of $H$ must transform under $SU(3)_L \times
SU(3)_R$ as $(1_L, \bar 3_R)$.  A suitable definition of parity is
thus$^{[13]}$
$$	P H_a (\vec x, t) P^{-1} = \gamma^0 H_b (-\vec x, t) \gamma^0
\Sigma_{ba} (- \vec x, t)\,\, . \eqno (8.19)$$

It is possible to redefine fields so that $H_a$ transforms in a simpler
way under parity.  Introduce
$$	\xi = \exp \left({iM\over f}\right) \,\, , \eqno (8.20)$$
which transforms under $SU(3)_L \times SU(3)_R$ as
$$	\xi \rightarrow L \xi U^{\dagger} = U \xi R^{\dagger} \,\, ,
\eqno (8.21)$$
since
$$	\xi^2 = \Sigma \,\, . \eqno (8.22)$$
Under parity
$$	P \xi (\vec x, t) P^{-1} = \xi^{\dagger} (- \vec x, t) \,\, .
\eqno (8.23)$$
In eq. (8.21) the special unitary matrix $U$ is (typically) a complicated
nonlinear function of
$L,R$ and the meson fields $M$.  Consequently $U$ depends on spacetime.
However, for elements of the unbroken $SU(3)_V$ subgroup $L = R = U$.
The redefined field
$$	\hat H_a = H_b \xi_{ba} \,\, , \eqno (8.24)$$
transforms under chiral $SU(3)_L \times SU(3)_R$ as
$$	\hat H_a \rightarrow \hat H_b U_{ba}^{\dagger}\,\, . \eqno (8.25)$$

The advantage of using the hatted fields is that the parity
transformation of eq. (8.19) becomes
$$	P\hat H_a (\vec x,t) P^{-1} = \gamma^0 \hat H_a (- \vec x,
t)\gamma^0 \,\, . \eqno (8.26)$$
For the remainder of this paper we shall use fields in the parity odd
$s_\ell = 1/2$ multiplet that transform under chiral symmetry as in eq. (8.25)
and under parity as in eq. (8.26), although for simplicity we shall not
put a hat on these fields.  Similar transformation laws hold for $G_a$ and
$F_a^\mu$.

Baryons with $Qq_a q_b$ flavor quantum numbers transforms according to the
$\bar
3$ and $6$ representations of $SU(3)_V$.  The $\bar 3$ contains an
isosinglet with zero strangeness ($\Lambda_Q$) and an isodoublet
with strangeness $-1 (\Xi_Q)$.  The lowest-lying heavy baryons in
the $\bar 3$ representation have $s_\ell = 0$ and positive parity.  We
denote the spin-1/2 fields that destroy these baryons by $T_a (T_3 =
\Lambda_Q, T_{1,2} = \Xi _Q)$  they transform under $SU(3)_L \times
SU(3)_R$ as
$$	T_a \rightarrow T_b U_{ba}^{\dagger} \,\, , \eqno (8.27)$$
and satisfy
$$	\vslash T_a = T_a \,\, . \eqno (8.28)$$
Under $SU(2)_v$ heavy quark spin symmetry $T_a \rightarrow ST_a$,
and under Lorentz transformations $T_a \rightarrow D
(\Lambda^{-1}) T_a$.

The lowest lying baryons in the 6 representation have
$s_\ell = 1$.  (The higher spin occurs because of fermi statistics.  The
ground state baryons in the six  have wave functions for the two light
quarks that are antisymmetric in color and
symmetric in flavor and space.  Therefore, they are symmetric in spin.)  This
angular momentum for the light degrees of freedom gives multiplets with
total spin $s_- = 1/2$ and $s_+ = 3/2$.  We denote the fields that destroy
these baryons by $S^{ab}$ and $S_\mu^{*ab}$ where $v^\mu S_\mu^{*ab} = 0$.
It is convenient to combine them into the object$^{[14]}$
$$	S_\mu^{ab} = \sqrt{{1\over 3}} (\gamma_\mu + v_\mu) \gamma^5
S^{ab} + S_\mu^{*ab} \,\, . \eqno (8.29)$$
Then under heavy quark spin symmetry
$$	S_\mu^{ab} \rightarrow S S_\mu^{ab} \,\, ,$$
where $S\epsilon SU(2)_v$.  Under Lorentz transformations $S_\mu^{ab}
\rightarrow \Lambda_{~~\nu}^\mu D(\Lambda^{-1}) S_\nu^{ab}$.  The combination
of fields $S_\mu^{ab}$ satisfies
$$	\vslash S_\mu^{ab} = S_\mu^{ab}, ~S_\mu^{ab} v^\mu = 0 \,\, .
\eqno (8.30)$$
\vfil\eject

\noindent {\bf 9.  Semileptonic $B \rightarrow De\bar \nu_e$ and $B
\rightarrow D^* e\bar \nu_e$ Decay}

Heavy quark spin-flavor symmetry determines many properties of hadrons
containing a single heavy quark.  Perhaps the best illustration of its
utility is provided by the semileptonic decays $B \rightarrow De\bar
\nu_e$ and $B \rightarrow D^* e\bar  \nu_e$.  In these decays the square
of the four-momentum transfer imparted by the virtual $W$-boson to the
heavy quarks is large,
$$	q^2 = (m_B v - m_D v')^2 = m_B^2 + m_D^2 - 2m_B m_D v \cdot v'
\,\, , \eqno (9.1)$$
and grows with the heavy quark masses.  However, as far as the light
degrees of freedom are concerned there isn't a large momentum transfer.
When the $B$ with four-velocity $v$ changes to a $D$ (or $D^*$) with
four-velocity $v'$ the light degrees of freedom go from a four-momentum
of order $\Lambda_{{\rm QCD}} v$ to a four-momentum of order
$\Lambda_{{\rm QCD}} v'$.  The square of the momentum transfer
felt by the light degrees of freedom is only of order
$$	q_\ell^2 \simeq (\Lambda_{{\rm QCD}} v - \Lambda_{{\rm QCD}}
v')^2 = - 2 \Lambda_{{\rm QCD}}^2 (v \cdot v' - 1)\,\, . \eqno (9.2)$$
Use of the effective heavy quark theory where $m_b$ and $ m_c \rightarrow
\infty$ is appropriate, because typical momentum transfers felt by the
light degrees of freedom are small compared with the heavy bottom and
charm quark masses.  Of course, there are always virtual gluons with
arbitrarily large momentum, and their effects are not adequately taken
into account by the effective theory.  Fortunately, because of
asymptotic freedom the differences between the full theory of  QCD and
the effective theory arising from high-momentum effects can be taken
into account using (renormalization-group improved) QCD perturbation
theory.  These high-momentum differences change the relationship between
the currents $\bar c \gamma_\mu \gamma_5 b$ and $\bar c\gamma_\mu b$ in QCD and
the operators that represent them in the effective theory.  In the
leading logarithmic approximation$^{[15]}$ (appropriate for $m_b \gg m_c \gg
\Lambda_{{\rm QCD}}$)
$$	\bar c \gamma_\mu (1 - \gamma_5)b = \left[{\alpha_s (m_b)\over
\alpha_s (m_c)}\right]^{-6/25} \left[{\alpha_s (m_c)\over\alpha_s
(\mu)}\right]^{a_{L}} \bar h_{v'}^{(c)} \gamma_\mu (1 -
\gamma_5)h_v^{(b)}\,\, , \eqno (9.3)$$
where
$$	a_L (v \cdot v') = {8\over 25} [v \cdot v' r (v \cdot v') -
1]\,\, , \eqno (9.4)$$
and
$$	r(v \cdot v') = {1\over \sqrt{(v \cdot v')^2-1}} \ln \left(v \cdot v'
+ \sqrt{(v \cdot v')^2 - 1}\right) \,\, . \eqno (9.5)$$

In the full theory the partially conserved current $\bar c \gamma_\mu (1
- \gamma_5)b$ doesn't require renormalization.  However, its matrix
elements contain (for $v \cdot v' \not= 1$) large logarithms of the
bottom and charm quark masses that become divergences in the effective
theory.  Consequently (for $v \cdot v'\not=1$) $h_{v'}^{(c)} \gamma_\mu
(1 - \gamma_5) h_v^{(b)}$ requires renormalization in the effective
theory and has dependence
on the subtraction point $\mu$.  This dependence cancels that of the
coefficient in eq. (9.3)

For $v = v'$ the current $\bar h_v^{(c)} \gamma_\mu (1 - \gamma_5)
h_v^{(b)}$ is not renormalized.  (This is consistent
with eq. (9.3) since $a_L (1) = 0$.) This is because  $\bar h_v^{(c)}
\gamma_\mu h_v^{(b)}$ is the conserved current associated with heavy
quark flavor symmetry.  Heavy quark spin symmetry ensures that
$\bar h_{v'}^{(c)} \gamma_\mu h_v^{(b)}$ is
renormalized in the same way as $\bar h_{v'}^{(c)} \Gamma h_v^{(b)}$,
where $\Gamma$ is any collection of gamma matrices.

For $B_a \rightarrow D_a$ and $B_a \rightarrow D_a^*$ matrix elements of
$\bar h_{v'}^{(c)} \gamma_\mu (1 - \gamma_5) h_v^{(b)}$, heavy quark spin
symmetry and  $SU(3)_V$ symmetry (the heavy quark
current is a singlet with respect to $SU(3)_V$) imply that$^{[9, 15]}$
$$	\bar h_{v'}^{(c)} \gamma_\mu (1 - \gamma_5) h_v^{(b)} = - \eta
(v \cdot v') Tr [\bar H_a^{(c)} (v') \gamma^\mu (1 - \gamma_5) H_a^{(b)}
(v)] \,\, , \eqno (9.6)$$
where $\eta$ is a universal function of $v \cdot v'$ independent of the
heavy quark masses, (such universal functions are commonly referred to as
Isgur-Wise functions).  $\eta$ has subtraction point dependence because
the current on the left-hand side of eq. (9.6) requires
renormalization in the effective heavy quark theory.

Note that heavy quark spin symmetry  forces $\gamma^\mu (1 -
\gamma_5)$ to occur between the $H$'s in eq. (9.6).  On the outside of
the $H$'s other factors like $\vslash$ or $\vslash'$ could occur, but
because of eq. (8.11) they can be reduced to the form in eq. (9.6).

Taking the traces in eq. (9.6) gives
$$	{<D(v')|\bar h_{v'}^{(c)} \gamma_\mu h_v^{(b)}| B(v)>\over
\sqrt{m_B m_D}} = \eta (v \cdot v') [v + v']_\mu \,\, , \eqno (9.7)$$

$$	{<D^* (v', \epsilon) |\bar h_{v'}^{(c)} \gamma_\mu \gamma_5
h_v^{(b)}|B(v)>\over\sqrt{m_B m_{D^{*}}}} = \eta (v \cdot v') [(1 + v \cdot
v') \epsilon_\mu^* - (\epsilon^* \cdot v) v'_\mu]\,\, , \eqno (9.8)$$

$$	{<D^* (v', \epsilon) |\bar h_{v'}^{(c)} \gamma_\mu h_v^{(b)} |
B(v)>\over \sqrt{m_B m_{D^{*}}}} = i \eta (v \cdot v')
\epsilon_{\mu\nu\alpha\beta} \epsilon^{*\nu} v^{'\alpha} v^\beta \,\,.
\eqno (9.9)$$
In eqs. (9.7), (9.8) and (9.9), the factors of $\sqrt{m_B m_D}$ and
$\sqrt{m_B m_{D^{*}}}$ are inserted
in the denominator because the heavy meson states $|M(\vec p,s)>$ are
normalized according to the usual convention
$$	<M(\vec p^{~\prime},s') |M(\vec p,s)> = 2 E \delta_{ss'} (2\pi)^3
\delta^3(\vec p - \vec p^{~\prime})\,\, ,$$
and the factor of the energy $E$ is proportional to the heavy meson
mass.

At zero recoil, $v = v'$, the vector current $\bar h_v^{(c)} \gamma_\mu
h_v^{(b)}$ is the conserved current associated with heavy quark flavor
symmetry.  Consequently its matrix element is fixed, implying that$^{[9,
16, 17]}$
$$	\eta (1) = 1\,\, . \eqno (9.10)$$

Eqs. (9.7)--(9.10) represent a tremendous amount of predictive power.
Lorentz (and parity) invariance imply that $B \rightarrow D$ and $B
\rightarrow D^*$ matrix elements of the vector and axial vector current
are parametrized by six form factors.  We have found that all those form
factors are simply related to the single function $\eta (v \cdot v')$
whose value at zero recoil is known.
\vfil\eject

\noindent {\bf 10.}  $\Lambda_{{\rm QCD}}/m_Q$ {\bf Corrections}

The Lagrange density for the heavy quark effective theory, given in eq.
(7.5), is valid for $m_Q \rightarrow \infty$.  The corrections that exist
at finite $m_Q$ can be found using a systematic expansion in powers of
$1/m_Q$.  The part of the QCD Lagrange density that involves a heavy
quark field $Q$ is
$$	{\cal L} = \bar Q (i\Dslash - m_Q)Q\,\, , \eqno (10.1)$$
and it implies the equation of motion
$$	(i\Dslash - m_Q) Q = 0\,\, . \eqno (10.2)$$
As shown in Section 7 to go over to the effective theory we write
$$	Q = e^{-im_{Q}v \cdot x} \left[h_v^{(Q)} + \chi_v^{(Q)}\right]
\,\, , \eqno (10.3)$$
where
$$	\vslash h_v^{(Q)} = h_v^{(Q)}, ~~ \vslash \chi_v^{(Q)} = -
\chi_v^{(Q)}\,\, . \eqno (10.4)$$
The field $\chi_v^{(Q)}$ does not represent the heavy antiquark.  It occurs
because the heavy quark is not precisely on-shell as it propagates.
Substituting (10.3) into (10.2) it is possible to solve for $\chi_v^{(Q)}$
order by order in $1/m_Q$.  Eq. (10.2) thus becomes
$$	[m_Q (\vslash - 1) + i\Dslash] [h_v^{(Q)} + \chi_v^{(Q)}] = 0
\,\, . \eqno (10.5)$$
Treating $\chi_v^{(Q)}$ as a small quantity we find
$$	\chi_v^{(Q)} = {1\over 2m_Q} i\Dslash h_v^{(Q)} + {\cal O}
(1/m_Q^2)\,\, . \eqno (10.6)$$
The derivative on $h_v^{(Q)}$ produces a factor of the residual momentum
which is typically of order $\Lambda_{{\rm QCD}}$.  Hence,
the expansion is in powers of $\Lambda_{{\rm QCD}}/m_Q$.  (By
$\Lambda_{{\rm QCD}}$ we mean a typical hadronic scale.  We do not
distinguish, in such order of magnitude estimates, between the chiral
symmetry breaking scale and the confinement scale.)  At
order $\Lambda_{{\rm QCD}}/m_Q$ the relationship between $Q$ and
$h_v^{(Q)}$ is
$$	Q = e^{-im_{Q} v \cdot x} \left[1 + {i\Dslash\over 2m_Q}\right]
h_v^{(Q)} \,\, .\eqno (10.7)$$
Putting this into (10.1) yields the Lagrange density$^{[18]}$
$$	{\cal L}_v = \bar h_v^{(Q)} i v \cdot D h_v^{(Q)} + {1\over
2m_Q} \bar h_v^{(Q)} \left[(iD)^2 - g_s \sigma_{\mu\nu}
G^{A\mu\nu}T^A\right] h_v^{(Q)} \,\, . \eqno (10.8)$$
The terms of order $\Lambda_{{\rm QCD}}/m_Q$ are to be treated as a
perturbation in the computation of $S$-matrix elements.  Hence the
equation of motion $v \cdot Dh_v^{(Q)}  = 0$ can be used to simplify the
couplings (e.g., $\bar h_v^{(Q)} (i v \cdot D)^2 h_v^{(Q)}$ vanishes
using the equation of motion).  Also note that the equation of motion
insures that the expression for $\chi_v^{(Q)}$ in eq. (10.6) is consistent
with the constraint $\vslash \chi_v^{(Q)} = - \chi_v^{(Q)}$.

In deriving the Lagrange density of eq. (10.8), we treated the gluon
field as a fixed background field and used the equation of motion
(10.2).  This amounts to matching tree graphs in the full theory of QCD
with tree graphs in the effective theory.  When quantum loop corrections
are included in the Lagrange density for the heavy quark effective
theory becomes
$$	{\cal L}_v = \bar h_v^{(Q)} i v \cdot D h_v^{(Q)} + {1\over
2m_Q} \bar h_v^{(Q)} \left[a_1 (iD)^2 - a_2 g_s\sigma_{\mu\nu}
G^{A\mu\nu} T^A\right]h_v^{(Q)}$$
$$	+ {\rm counter~ terms}\,\, . \eqno (10.9)$$
The couplings $a_1$ and $a_2$ are subtraction-point dependent.  The tree
level matching in eq. (10.8) determines that
$$	a_1 (m_Q) = 1 + {\cal O} (\alpha_s (m_Q)) \,\, , \eqno (10.10a)$$
$$	a_2 (m_Q) = 1 + {\cal O} (\alpha_s (m_Q)) \,\, . \eqno (10.10b)$$
The $\mu$ dependence of $a_{1,2}$ follows from the renormalization of
the operators $\bar h_v^{(Q)} (iD)^2 h_v^{(Q)}$ and $\bar h_v^{(Q)} g_s
\sigma_{\mu\nu}G^{A\mu\nu} T^A h_v^{(Q)}$.  Explicit calculation
demonstrates that (in the
leading logarithmic approximation) $a_1$ is independent of $\mu$ and
$$	a_2 (\mu) = \left[{\alpha_s (m_Q)\over\alpha_s
(\mu)}\right]^{-9/(33-2N)}\,\, , \eqno (10.11)$$
where $N$ is the number of quark flavors appropriate to the momentum
interval between $m_Q$ and $\mu$.

The fact that $a_1 (\mu) = 1$ can be understood as a consequence of
reparametrization invariance.$^{[19]}$  Recall that the  heavy quark
four-momentum is the sum of a term proportional to $m_Q$ and the
residual momentum.
$$	p = m_Q v + k \,\, . \eqno (10.12)$$
However, this decomposition is not unique.  The physics must be the same
if the small changes
$$	v \rightarrow v + \epsilon/m_Q \,\, , \eqno (10.13a)$$
$$	k \rightarrow k - \epsilon \,\, , \eqno (10.13b)$$
are made.  Since the four-volicity satisfies $v^2 = 1$ the infinitesimal
parameter $\epsilon$ satisfies
$$	v \cdot \epsilon = 0 \,\, . \eqno (10.14)$$
In addition to the changes in eq. (10.13) to preserve the constraint
$\vslash h_v^{(Q)} = h_v^{(Q)}$ the heavy quark field changes,
$h_v^{(Q)} \rightarrow h_v^{(Q)} + \delta h_v^{(Q)}$, where
$$	\delta h_v^{(Q)} = {\eslash\over 2m_Q} h_v^{(Q)} \,\, . \eqno
(10.15)$$
Note that $\vslash \delta h_v^{(Q)} = - h_v^{(Q)}$.  Neglecting, for the
moment, the gauge fields, the Lagrange density (10.9) should be invariant
under (10.13).  Since a derivative brings down a factor of the residual
momentum, replacing $D_\mu \rightarrow - i k_\mu$ in eq. (10.9) and then
demanding invariance under (10.13) gives $a_1 (\mu) = 1$.  Because it
follows from reparametrization invariance this result holds to all
orders in perturbation theory.

In eq. (10.9) only the last term violates heavy quark
spin symmetry.  It is the matrix element of this term that gives rise to
the $P_Q^* - P_Q$ mass difference.  Since it is the unique operator, $\bar
h_v^{(Q)} g_s\sigma_{\mu\nu} G^{A\mu\nu} T^A h_v^{(Q)}$, that causes the
splitting for two heavy quarks$^{[20]}$  $Q_i$ and $Q_j$, we deduce
$$	m_{P_{Q_{i}}^{*}} - m_{P_{Q_{i}}} = \left({m_{Q_{j}}\over
m_{Q_{i}}}\right) \left[{\alpha_s (m_{Q_{i}})\over\alpha_s
(m_{Q_{j}})}\right]^{-9/(33-2N)} (m_{P_{Q_{j}}^{*}} - m_{P_{Q_{j}}}) \,\,
, \eqno (10.16)$$
where $N$ denotes the number of quark flavors appropriate to the
momentum interval between $m_{Q_{i}}$ and $m_{Q_{j}}$.  The measured
$B^* - B$ and $D^* - D$ mass differences agree well with eq. (10.12).

\noindent {\bf 11.  Chiral Lagrangian for Heavy Mesons}

The ground state heavy mesons have $s_\ell^{\pi_{\ell}} = 1/2^-$ for the
spin-parity of the light degrees of freedom.  The low-momentum strong
interactions of these heavy mesons are described by a Lagrange density
that is invariant under chiral $SU(3)_L \times SU(3)_R$ symmetry, heavy
quark symmetry, parity, and Lorentz transformations.  (For invariance under
parity and Lorentz transformations explicit factors of $v$ are treated as if
the four-velocity transforms
are a true four-vector.)  The chiral Lagrange density that describes the
low momentum strong interactions of
heavy $P_a$ and $P_a^*$ mesons is$^{[21, 22, 23]}$
$$	{\cal L} = - i Tr \bar H_a v_\mu \partial^\mu H_a + {i\over 2}
Tr \bar H_a H_b v^\mu (\xi^{\dagger} \partial_\mu \xi + \xi
\partial_\mu \xi^{\dagger})_{ba}$$
$$	+ {i\over 2} g Tr \bar H_a H_b \gamma_\nu \gamma_5
(\xi^{\dagger} \partial^\nu \xi - \xi \partial^\nu \xi^{\dagger})_{ba} +
... \eqno (11.1)$$
Here $H_a$ represents the $4 \times 4$ matrix containing the heavy
meson fields defined in eq. (8.6), $\xi$ contains the Goldstone boson
$\pi, K$ and $\eta$ fields as defined in eq. (8.20), and the ellipsis
denotes terms with more derivatives.  The traces in eq. (11.1) are over
the $4\times 4$ matrices.  The SU(3) indices $a,b$ are explicitly
displayed and repeated indices are summed over 1,2,3.  Factors of
$\sqrt{m_{P_{a}}}$ and $\sqrt{m_{P^{*}_{a}}}$ have been absorbed into
the heavy meson $P_a$ and $P_a^*$ fields so that they have dimension 3/2.

Eq. (11.1) has been simplified using eqs. (8.11).  For example, (8.11)
implies that the term $Tr \bar H_a H_b \gamma_5 v_\nu (\xi^{\dagger}
\partial^\nu \xi - \xi \partial^\nu \xi^{\dagger})_{ba}$ vanishes.  In
the terms of eq. (11.1) no gamma matrices can occur between the factors
of $\bar H$ and $H$ because of heavy quark spin symmetry.

Expanding $\xi$ in powers of the Goldstone boson matrix $M$ and taking
the traces yields Feynman rules for the interaction of $\pi, K$, and
$\eta$ with the heavy mesons.  The $P_a$ and $P^*_a$ propagators that
follow from eq. (11.1) are $i\delta_{ab}/2v \cdot k$ and $-i \delta_{ab}
(g_{\mu\nu} - v_\mu v_\nu)/2v\cdot k$ respectively.  The $P P^* M$ and
$P^* P^* M$ couplings arise from the term proportional to $g$ in the
Lagrange density (11.1).  The second term does not give rise to
couplings of the heavy mesons to a single Goldstone boson field.  Heavy
quark flavor symmetry implies that the coupling $g$ is independent of
the heavy quark mass, $m_Q$.

The light quark mass terms in the QCD Lagrange density transform as
$(\bar 3_L, 3_R) + (3_L, \bar 3_R)$ under chiral $SU(3)_L \times
SU(3)_R$.  To incorporate the leading effects of explicit symmetry
breaking from light quark masses, we add
$$	\eqalign{\delta {\cal L}^{(1)} &= \lambda_1 Tr \bar H_b H_a (\xi m_q \xi +
\xi^{\dagger} m_q \xi^{\dagger})_{ab}\cr
	&+ \lambda'_1 Tr \bar H_a H_a (\xi m_q \xi + \xi^{\dagger} m_q
\xi^{\dagger})_{bb} + ...\cr} \eqno (11.2)$$
to the chiral Lagrange density.  The ellipsis denotes
terms with more derivatives and $m_q$ represents the light quark mass matrix
(see eq. (2.21)).  The first term in eq. (11.2) fixes the mass splitting
between heavy mesons containing an anti-strange quark and  those
containing anti-up or anti-down quarks.  The second term contributes an
equal amount to the heavy meson masses and does not contribute to  SU(3)
violating heavy meson mass splittings.  The couplings $\lambda_1$ and
$\lambda'_1$ are independent of the heavy quark mass (in the $m_Q
\rightarrow \infty$ limit).

It is also possible to include deviations from the $m_Q \rightarrow
\infty$ limit that violate heavy quark symmetry.  At order $\Lambda_{\rm
QCD}/m_Q$ the heavy quark spin symmetry is broken only by the color
magnetic moment operator $\bar h_v^{(Q)} g_s \sigma_{\mu\nu} G^{A \mu\nu}
T^A h_v^{(Q)}$.  This operator is a singlet under chiral $SU(3)_L \times
SU(3)_R$, and to include its effects
$$	\delta {\cal L}^{(2)} = {\lambda_2\over m_Q} Tr \bar H_a
\sigma^{\mu\nu} H_a \sigma_{\mu\nu} + ... \eqno (11.3)$$
is added to the Lagrange density.  Here the ellipsis denotes terms with
derivatives.  It follows from eq. (10.11) that the coupling $\lambda_2$
has logarithmic dependence on $m_Q$.  The term in eq. (11.3)  is
independent of the Goldstone boson fields.  Its only effect is to shift
the $P^*$ and $P$ masses.  In terms of the mass difference
$$	\Delta = m_{P^{*}} - m_P = - {8 \lambda_2\over m_Q}\,\, , \eqno
(11.4)$$
the propagators for the $P_a$ and $P_a^*$  mesons become $i\delta_{ab}/2
(v \cdot k + {3\over 4} \Delta)$ and $-i\delta_{ab} (g_{\mu\nu} - v_\mu
v_\nu)/2(v\cdot k - {1\over 4} \Delta)$.  Now in the rest frame $v = (1,\vec
0)$
an on-shell $P_a$ meson has residual energy $- {3\over 4} \Delta$ and an
on-shell $P_a^*$ meson has residual energy ${1\over 4}\Delta$.  It is
convenient when dealing with situations involving a real $P_a$ meson and
a virtual $P_a^*$ meson to redefine the heavy meson fields by $\exp
({i3\over 4} \Delta v \cdot x)$ so the $P_a$ and $P_a^*$ propagators become
$i\delta_{ab}/2 v \cdot k$ and $- i\delta_{ab} (g_{\mu\nu} - v_\mu
v_\nu)/2(v \cdot k - \Delta)$ respectively.

The $\Lambda_{\rm QCD}/m_Q$
corrections due to the operator $\bar h_v^{(Q)} (iD)^2 h_v^{(Q)}$
violate heavy quark flavor symmetry and cause the couplings $g,
\lambda_1$ and $\lambda'_1$ to depend on the heavy quark mass $m_Q$.

In the next few sections we explore the implications of chiral
perturbation theory for the interactions of heavy mesons.  The
combination of chiral and heavy quark symmetry provides a powerful tool
for studying these interactions.

\noindent {\bf 12.  The Coupling} $g$

In chiral perturbation theory for heavy mesons, the fundamental coupling
is $g$.  The heavy meson contribution to the light quark axial current
is obtained from the Lagrange density in eq. (11.1) using the Noether
procedure.  Under an infinitesimal axial transformation $\delta M = -
f \epsilon^A T^A + ...$, while eqs. (8.18) and (8.24) imply that $H_a
\rightarrow H_a + ...$~~.  Here the ellipsis denotes terms containing the
Goldstone boson fields $M$.  It follows that
$$	\bar q_a T_{ab}^A \gamma_\nu \gamma_5 q_b = - g Tr \bar H_a H_b
\gamma_\nu \gamma_5 T_{ab}^A + ... \,\, . \eqno (12.1)$$
Treating the quark fields in eq. (12.1) as constituent quarks
and using the nonrelativistic constituent quark model to estimate the
$D^*$ matrix element of left hand side of eq. (12.1) gives $g = 1$.  A
similar estimate of the pion nucleon coupling implies that $g_A = 5/3$.
(Recall that experimentally $g_A = 1.25$.)  Thus our expectation is that $g$
is around unity.

Expanding the Lagrangian (11.1) in powers of the Goldstone boson fields
$M$, we find at linear order that

$$	\eqalign{{\cal L} &= {-g\over f} Tr \bar H_a H_b \gamma_\nu
\gamma_5 \partial^\nu M_{ba}\cr
\cr
 	& = \left[\left({-2g\over f}\right) \partial^\nu M_{ba}
P_a^{\dagger} P_{b\nu}^* +h.c.\right] + \left({2gi\over f}\right)
\partial^\nu M_{ba} P_a^{*\alpha\dagger} P_b^{*\beta}
\epsilon_{\alpha\lambda\beta\nu} v^\lambda \,\, .\cr} \eqno (12.2)$$
Eq. (12.2) contains the $P^* PM$ and $P^* P^* M$ couplings.  (Note that
because of parity invariance there is no PPM coupling.)  Using eq. (12.2)
for $Q = c$ and $M = \pi$ gives

$$	\Gamma (D^{*+} \rightarrow D^0 \pi^+) = {g^2\over 6\pi f^2}
|\vec p_\pi |^3 \,\, . \eqno (12.3)$$
The decay width for $D^{*+} \rightarrow D^+ \pi^0 $ is a factor of two
smaller by isospin symmetry.  The experimental upper limit$^{[24]}$
of 131 KeV on
the $D^*$ width when combined with the $D^{*+} \rightarrow D^+ \pi^0$
and $D^{*+} \rightarrow D^0 \pi^+$ branching ratios$^{[25]}$
of Table I imply that
$g^2 \lsim 0.5$.  In evaluating eq. (12.3), $f = f_\pi \simeq 132$ MeV
was used.
\vfil\eject

\centerline{TABLE I}
\input tables
\bigskip
\begintable {\bf Decay Mode} | {\bf Branching Ratio \%}~~\crthick
$D^{*0} \rightarrow D^0 \pi^0$ | $63.6 \pm 2.3 \pm 3.3$\crnorule
$D^{*0} \rightarrow D^0 \gamma~$ | $36.4 \pm 2.3 \pm 3.3$\cr
$D^{*+} \rightarrow D^0 \pi^+$ | $68.1 \pm 1.0 \pm 1.3$\crnorule
$D^{*+} \rightarrow D^+ \pi^0$ | $30.8 \pm 0.4 \pm 0.8$\crnorule
$D^{*+} \rightarrow D^+ \gamma~$ | $~1.1 \pm 1.4 \pm 1.6$\endtable

Even if the $D^*$ decay width is too small to measure, radiative $D^*$
decay may provide a (indirect) determination of $g.^{[26,27]}$  The $D_a^*
\rightarrow D_a \gamma$ matrix element has the form
$$	{\cal M} (D_a^* \rightarrow D_a \gamma) = e \mu_a
\epsilon^{\mu\alpha\beta\lambda} \epsilon_\mu^* (\gamma) v_\alpha k_\beta
\epsilon_\lambda (D^*) \,\, , \eqno (12.4)$$
where $e\mu_a/2$ is the transition magnetic moment, $k$ is the photon
momentum, $\epsilon(\gamma)$ is the polarization vector for the photon, and
$\epsilon (D^*)$ is the polarization vector for the $D^*$.  The
resulting decay rate is
$$	\Gamma (D_a^* \rightarrow D_a \gamma) = {\alpha_e\over 3}
|\mu_a|^2 |\vec k|^3 \,\, . \eqno (12.5)$$
The $D_a^* \rightarrow D_a \gamma$ matrix element gets contributions
from the photon coupling to the light quark part of the electromagnetic
current, ${2\over 3} \bar u \gamma_\mu u - {1\over 3} \bar d \gamma_\mu
d - {1\over 3} \bar s \gamma_\mu s$, and the photon coupling to the
heavy charm quark part of the electromagnetic current, ${2\over 3} \bar
c \gamma_\mu c$.  The part of $\mu_a$ that comes from the heavy charm
quark piece of the electromagnetic current, $\mu^{(h)}$, is determined
by heavy quark symmetry.  In the effective heavy quark theory the
Lagrange density for strong and electromagnetic interactions of the
charm quark is
$$	{\cal L}_\nu = \bar h_v^{(c)} (i v \cdot D) h_v^{(c)} + {1\over
2m_c} \bar h_v^{(c)} (iD)^2 h_v^{(c)}$$
$$	- {g_s\over 2m_c} \bar h_v^{(c)} \sigma^{\mu\nu} T^A h_v^{(c)}
G_{\mu\nu}^A - {e\over 3m_c} \bar h_v^{(c)} \sigma^{\mu\nu} h_v^{(c)}
F_{\mu\nu} +... \,\, , \eqno (12.6)$$
where the ellipsis denotes terms suppressed by more factors of
$1/m_c$, and the subtraction point is chosen to be $\mu = m_c$.   This is
an extension of the result presented in eq. (10.9) to include
electromagnetic interactions,
$$	D_\mu = \partial_\mu + ig_s A_\mu^A T^A + {2\over 3} ie A_\mu
\,\, . \eqno (12.7)$$
Only the last term in eq. (12.6) contributes to the $D_a^* \rightarrow
D^* \gamma$ matrix element.  Using the generalization of eq. (9.6)
$$	\bar h_{v'}^{(c)} \Gamma h_v^{(c)} = - \eta (v \cdot v') Tr
[\bar H_a^{(c)} (v') \Gamma H_a^{(c)} (v)] \,\, , \eqno (12.8)$$
with $\Gamma = \sigma^{\mu\nu}$ and $v' = v$, implies (using $\eta(1) =
1$) that
$$	\mu^{(h)} = {2\over 3m_c} \,\, . \eqno (12.9)$$

The part of $\mu_a$ that comes from the photon coupling to the light
quark piece of the electromagnetic current, $\mu_a^{(\ell)}$, is not fixed
by heavy quark symmetry.  The light quark piece of the electromagnetic
current transforms as an octet under the unbroken $SU(3)_V$ flavor
symmetry group.  Since there is only one way to combine an 8,3, and $\bar
3$ into a singlet, the $\mu_a^{(\ell)}$ are expressible in terms of a
single reduced matrix element
$$	\mu_a^{(\ell)} = Q_a \beta \,\, , \eqno (12.10)$$
where $\beta$ is an unknown constant and $Q_a$ denotes the light quark
charges $Q_1 = 2/3, ~~Q_2 = - 1/3, ~~Q_3 = - 1/3$.  (In the nonrelativistic
constituent quark model, $\mu_a^{(\ell)}$ arises from the magnetic
moment of a constituent quark.  This leads to the expectation that
$\beta \simeq 3 GeV^{-1}$.)  Eq. (12.10) includes effects suppressed by
powers of $1/m_c$ since it follows only from  $SU(3)_V$ symmetry.

The leading $SU(3)_V$ violating contribution to $\mu_a$ has a
nonanalytic dependence on the light quark masses $m_q$ of the form
$m_q^{1/2}$, and arises from the one-loop Feynman diagrams in Figure 3.
It is straightforward to compute these diagrams.  Gauging the Goldstone
boson chiral Lagrange density in eq. (3.11),
$$	{\cal L} = {f^2\over 8} (D_\mu \Sigma) (D_\mu \Sigma)^{\dagger} \,\, ,
\eqno (12.11)$$
where
$$	D_\mu \Sigma = \partial_\mu \Sigma + ie [Q, \Sigma] {\cal
A}_\mu\,\, , \eqno (12.12)$$
and
$$	Q = \pmatrix{2/3 & 0 & 0\cr
	0 & - 1/3 & 0 \cr
	0 & 0 & - 1/3\cr}\,\, . \eqno (12.13)$$
 Eq. (12.11) gives the electromagnetic interactions of the Goldstone
bosons,  and it yields the photon vertices in Figure 3.  The remaining
vertices needed for Figure 3 follow from eq. (12.2).  The resulting
loop integral has the form
$$	\eqalign{I^{\nu\alpha} & \equiv \int {d^n q\over (2\pi)^n}
{q^\nu q^\alpha\over (v \cdot q + i\epsilon) (q^2 - m_K^2 +
i\epsilon)^2}\cr
	& = 4 \int_0^\infty d\lambda \int {d^n q\over (2\pi)^n} {q^\nu
q^\alpha\over (q^2 + 2\lambda v \cdot q - m_K^2 + i\epsilon)^3}\cr
	& = \left({4\over n}\right) g^{\nu\alpha} \int_0^\infty d\lambda
\int {d^n q\over (2\pi)^n} {q^2\over (q^2 - m_K^2 - \lambda^2 +
i\epsilon)^3} + ... \,\, ,\cr} \eqno (12.14)$$
where the ellipsis denotes terms proportional to $v^\nu v^\alpha$ that
don't contribute to the amplitude.  Finally the $q$ and $\lambda$
integrations are performed using the formulas
$$	\int {d^n q\over (2\pi)^n} {(q^2)^\alpha\over (q^2 -
M^2 + i\epsilon)^\beta} = i (-1)^{\alpha + \beta} (M^2)^{{n\over 2} +
\alpha - \beta} {\Gamma (\alpha + n/2) \Gamma (\beta - \alpha - n/2)\over
\Gamma (n/2) \Gamma (\beta)} \eqno (12.15)$$
and
$$	\int_0^\infty d\lambda (1 + \lambda^2)^{-p} = {\pi \Gamma (2p -
2)\over 2^{2p-2} \Gamma (p-1) \Gamma (p)} \,\,\, . \eqno (12.16)$$
This gives as $n \rightarrow 4$,
$$	I^{\nu\alpha} = {-i\over 16\pi} g^{\nu\alpha} m_K +... \,\, . \eqno
(12.17)$$
where the ellipsis denotes terms proportional to $v^\nu v^\alpha$.

Including the $SU(3)_V$ violations that follow from Figure 3 the
expression for $\mu_a^{(\ell)}$ becomes
$$	\mu_1^{(\ell)} = {2\over 3}\beta - {g^2 m_K\over 4\pi
f^2} - {g^2 m_\pi\over 4\pi f^2} 	\eqno (12.18a)$$
$$	\mu_2^{(\ell)} = {-1\over 3} \beta + {g^2 m_\pi\over 4\pi f^2}
\eqno (12.18b)$$
$$	\mu_3^{(\ell)} = {-1\over 3} \beta + {g^2 m_K\over 4\pi f^2}
\,\, . \eqno (12.18c)$$
For $m_K \not= m_\pi$ the one-loop contribution to $\mu_1^{(\ell)},
\mu_2^{(\ell)}$ and $\mu_3^{(\ell)}$ is not in the ratio $2: -1: -1$ and
hence violates $SU(3)_V$.  The most important corrections to eqs.
(12.18) come from $SU(3)_V$ violating terms of order $m_s$.  These are
analytic in the strange quark mass and so are not calculable.

Using
$$	\mu_a = \mu_a^{(\ell)} + \mu ^{(h)} \,\,\, , \eqno (12.19)$$
with $\mu_a^{(\ell)}$ and $\mu^{(h)}$ given by eqs. (12.18) and (12.9),
determines the rates for $D^{0*} \rightarrow D^0 \gamma, D^{*+}
\rightarrow D^+\gamma$ and $D_s^* \rightarrow D_s \gamma$ in terms of
$\beta$ and $g$.  The measured ratio of branching ratios $Br(D^{*0}
\rightarrow D^0 \gamma)/ Br (D^{*0} \rightarrow D^0 \pi^0)$ thus gives
$\beta$ as a function of $g$.  The ratio of branching functions
$Br (D^{*+} \rightarrow D^+ \gamma)/Br (D^{*+} \rightarrow D^0 \pi^+)$
can therefore be expressed in terms of $g$.  At the present time there is only
an upper limit
on the value of the $Br(D^{*+} \rightarrow D^+ \gamma)$, but when it is
measured a value for the coupling of $g$ can be extracted.
\vfil\eject

\noindent {\bf 13. ~Semileptonic $B \rightarrow De\bar\nu_e$ and
$B\rightarrow D^*e\bar\nu_e$ Decay at Zero Recoil}

At zero recoil $({\rm i.e}., v = v')$ heavy quark symmetry implies that
$$	\eqalignno{{<D(v) |\bar c \gamma_\mu b|B(v)>\over \sqrt{m_B
m_D}} & = \left[{\alpha_s (m_b)\over\alpha_s (m_c)}\right]^{-6/25}
2v_\mu \,\, ,& (13.1a)\cr
	{<D^*(v,\epsilon)|\bar c \gamma_\mu \gamma_5
b|B(v)>\over\sqrt{m_B m_{D^{*}}}} &= \left[{\alpha_s
(m_b)\over\alpha_s(m_c)}\right]^{-6/25} 2\epsilon_\mu^*\,\, . & (13.1b)\cr}$$
Furthermore, it has been shown that there are no order $1/m_c$ or $1/m_b$
corrections to the relations (13.1).  Chiral perturbation theory has
been used to examine the order $(1/m_c)^{n+2}, n = 0,1,2,...$
corrections to eqs. (13.1).  For small up and down quark masses, the
leading corrections result from the one-loop diagram in Figure 4 and
wavefunction renormalization.  In Figure 4 the shaded square denotes
an insertion of the weak current vertices (13.1), and the shaded circles
denote $P^* P^*\pi$ or $P^* P\pi$ vertices (see eq. (12.2)) from the
chiral Lagrangian in eq. (11.1).  This gives$^{[28]}$
$$	\eqalignno{{<D(v) | \bar c \gamma_\mu b|B(v)>\over\sqrt{m_B
m_D}} &= \left[{\alpha_s (m_b)\over\alpha_s (m_c)}\right]^{-6/25} 2v_\mu
\Bigg\{1 + C(\mu) - {3g^2\over 2} \left({\Delta^{(c)}\over 4\pi
f}\right)^2\cr
	&~~~\cdot \left[\ell n (\mu^2/m_\pi^2) + f
(\Delta^{(c)}/m_\pi)\right]\Bigg\} & (13.2a)\cr
	{<D^*(v,\epsilon)|\bar c \gamma_\mu \gamma_5 b|
B(v)>\over\sqrt{m_B m_{D^{*}}}} &= \left[{\alpha_s (m_b)\over\alpha_s
(m_c)}\right]^{-6/25} 2\epsilon_\mu^* \Bigg\{1 + C' (\mu) -{g^2\over 2}
\left({\Delta^{(c)}\over 4\pi f}\right)^2\cr
	&~~~\cdot \left[\ell n (\mu^2/m_\pi^2) +
f(-\Delta^{(c)}/m_\pi)\right]\Bigg\} \,\, , & (13.2b)\cr}$$
where
$$ 	f(x) = 2 \int_0^\infty dq {q^4\over (q^2 + 1)^{3/2}}
\left\{{1\over [(q^2 + 1)^{1/2} + x]^2} - {1\over q^2 + 1}\right\} \,\,
. \eqno (13.3)$$

In eqs. (13.2) $C(\mu)$ and $C'(\mu)$ are the contribution of tree level
``counter terms'' of order $1/m_c^2$, and $\Delta^{(c)} = m_{D^{*}} -
m_D$.  The dependence of $C$ and $C'$ on the subtraction point $\mu$ is
cancelled by that of the logarithm.  For $\mu$ of order the chiral
symmetry breaking scale, $\sim 1 GeV, C(\mu)$ and $C'(\mu)$ contain no
large logarithms and (at least formally) are less important than the
terms with a logarithm of the pion mass (which are also of order
$1/m_c^2$ since $\Delta^{(c)}$ is order $1/m_c$).  The function $f$
takes into account the effects of corrections of order $(1/m_c)^{2+n}, n
= 1,2,...$~.  It is enhanced by powers of $1/m_\pi$ over terms we have
neglected and should provide a reliable estimate of the order
$(1/m_c)^{2+n}, n = 1,2,...$ effects.  Because
the pion mass occurs in the denominator, the expansion in powers of
$1/m_c$ breaks down in the limit where the pion mass goes to zero.
Experimentally, $m_\pi$ is about equal to $\Delta^{(c)}$ and so all the terms
of order $(1/m_c)^{2+n}, n = 1,2,3,...$ are of comparable importance.

Since $\Delta^{(c)}$ is greater than $m_\pi$ the $B \rightarrow D^*$
matrix element has an imaginary part.  However, experimentally
$\Delta^{(c)}$ is very close to $m_\pi$ and it is a good approximation
to set $\Delta^{(c)}/m_\pi = 1$.  The expression in eq. (13.3) gives
$f(1) = 2 ({7\over 3} - \pi)$ and $f(-1) = 2({7\over 3} + \pi)$.
Numerically for $g^2 = 0.5$ and $\mu = 1$ GeV, the correction to the $B
\rightarrow D$ matrix element from the ``large logarithm'' is -2.1\% and
the correction from $f$ is 0.9\%.  For the $B \rightarrow D^*$ matrix
element the correction from the large logarithm is -0.7\% and from $f$
is -2.0\%.

In this section we have only used chiral $SU(2)_L \times SU(2)_R$.  The
order $1/m_c^2$ effects of kaon and eta loops are absorbed into the
constants $C(\mu)$ and $C'(\mu)$.

\noindent {\bf 14. ~Semileptonic $B \rightarrow \pi e\bar\nu_e$ or $D
\rightarrow \pi\bar e\nu_e$ Decay}

For most of the Dalitz plot, chiral perturbation theory cannot be applied
to $B \rightarrow \pi e\bar\nu_e$ and $D \rightarrow \pi\bar e\nu_e$
decay since (in the $B$ or $D$ rest frame) the pion has a large energy compared
with the chiral
symmetry breaking scale.  In this section we focus on the tiny region of
phase space where the pion has an energy small enough that chiral
perturbation theory can be applied.  For definiteness let's focus on the
decay  $B \rightarrow \pi e\bar\nu_e$.  Then the relevant hadronic matrix
element is
$$	<\pi (p_\pi)|\bar u \gamma_\mu (1 - \gamma_5)b|B(v)> =
f_+ (p_B + p_\pi)_\mu +f_- (p_B - p_\pi)_\mu \,\, , \eqno (14.1)$$
where $p_B = m_B v$.

The semileptonic decays $B \rightarrow \pi e\bar\nu_e$ and $B
\rightarrow \pi\mu\bar\nu_\mu$ depend only on $f_+$;  the contribution
of $f_-$ is proportional to the lepton mass and can be neglected.  In
the large $b-$quark mass limit (when $v\cdot p_\pi\ll m_b$), the
left-hand side
goes as $\sqrt{m_b}$ from the normalization of states (there is also
a logarithmic dependence on $m_b$ from perturbative QCD effects).
Consequently, for $v \cdot p_\pi \ll m_b$,
$$	f_+ + f_- \sim {\cal O} (1/\sqrt{m_b}) \,\, , \eqno (14.2a)$$
$$	f_+ - f_- \sim {\cal O} (\sqrt{m_b}) \,\, . \eqno (14.2b)$$
So, in the limit $m_b \rightarrow \infty$, $f_+ = - f_-$.  The known dependence
of
the form factors $f_\pm$ on the heavy quark mass (and isospin symmetry)
means that form factors for $B \rightarrow \pi e\bar\nu_e$ are related
to those for $D \rightarrow \pi \bar e \nu_e$.  Including perturbative
QCD effects, we find the relationship$^{[29]}$
$$	(f_+^{(B \rightarrow \pi)} + f_-^{(B \rightarrow
\pi)}) = \sqrt{{m_D\over m_B}} \left[{\alpha_s (m_b)\over\alpha_s
(m_c)}\right]^{-6/25} (f_+^{(D \rightarrow \pi)} + f_-^{(D \rightarrow
\pi)}) \,\, , \eqno (14.3a)$$
$$	(f_+^{(B \rightarrow \pi)} - f_-^{(B \rightarrow \pi)}) =
\sqrt{{m_B\over m_D}} \left[{\alpha_s (m_b)\over\alpha_s
(m_c)}\right]^{-6/25} (f_+^{(D \rightarrow \pi)} - f_-^{(D \rightarrow
\pi)}) \,\, . \eqno (14.3b)$$
Since $f_+ = - f_-$  as $m_Q \rightarrow \infty, $ eq. (14.3b) implies the
important relation
$$	f_+^{(B\rightarrow \pi)} = \sqrt{{m_B\over m_D}} \left[{\alpha_s
(m_b)\over\alpha_s (m_c)}\right]^{-6/25} f_+^{(D \rightarrow \pi)}\,\, .
\eqno (14.4)$$
Eqs. (14.3) and (14.4) are consequences of heavy quark flavor symmetry.
Naively, these formulae hold as long as $v \cdot p_\pi$ is small
compared to the heavy quark masses $m_c$ and $m_b$.  However, we shall
see shortly that, for very small $v \cdot p_\pi$, chiral perturbation
theory implies that eqs. (14.3) and (14.4) are not valid.$^{[30]}$

The operator
$$	L_a^\nu = \bar q_a \gamma^\nu (1 - \gamma_5) b \,\, , \eqno (14.5)$$
transforms under chiral $SU(3)_L \times SU(3)_R$ as $(\bar 3_L, 1_R)$,
and in chiral perturbation theory its hadronic matrix elements are given
by those of
$$	L_a^\nu = \left({i\alpha\over 2}\right) Tr \gamma^\nu (1 -
\gamma_5)H_b \xi_{ba}^{\dagger} +... \,\, , \eqno (14.6)$$
where the ellipsis denotes terms with derivatives, factors of the light
quark mass matrix $m_q$, or factors of $1/m_Q$.  The constant $\alpha$
has a logarithmic dependence on the $b-$quark mass and is related to the
$B$ meson decay constant $f_B$.  Using equation (14.6) with $a = 1$ to
calculate the matrix element
$$	<0|\bar u \gamma_\nu \gamma_5 b|B^- (v)> = if_B p_B^\nu\,\, ,
\eqno (14.7)$$
gives
$$	\alpha = f_B \sqrt{m_B}\,\, . \eqno (14.8)$$
The form factors $f_\pm$ are given by the $B \rightarrow \pi$ matrix
element of $L_1^\nu$ in eq. (14.6).  Calculating the Feynman diagrams in
Figure 5 and using eq. (14.8) to express $\alpha$ in
terms of $f_B$ gives$^{[21, 22, 23]}$ for $B^o \rightarrow \pi^+
e\bar\nu_e$
$$	\eqalignno{f_+ + f_- &= - (f_B/f) [1 - g v \cdot p_\pi/(v \cdot
p_\pi + \Delta^{(b)})]\,\, , & (14.9a)\cr
	f_+ - f_- &= - g f_B m_B/f (v \cdot p_\pi + \Delta^{(b)})\,\, , &
	 (14.9b)\cr}$$
where $\Delta^{(b)} = m_{B^{*}} - m_B$.  For $B^- \rightarrow \pi^o e
\bar\nu_e$ decay there is an additional $1/\sqrt{2}$.  Eqs. (14.9) are valid
for $v
\cdot p_\pi$ much less than the chiral symmetry breaking scale.  Note
that they don't depend on heavy quark flavor symmetry but do use the
heavy quark spin symmetry.  (In the pole graph it is the heavy quark spin
symmetry that relates the $B^*$ decay constant to that of the $B$.)
Eqs. (14.9) indicate that $f_+ + f_-$ is negligible compared with $f_+ - f_-$
provided $g$ is not too small.  For $g$ around unity
$$	f_+ = - g f_B m_B/2f(v \cdot p_\pi + \Delta^{(b)})\,\, . \eqno
(14.10)$$
Eqs. (14.9) and (14.10) also hold for $D \rightarrow \pi$ provided one
replaces $f_B \rightarrow f_D, m_B \rightarrow m_D$ and $\Delta^{(b)}
\rightarrow \Delta^{(c)} = m_{D^{*}} - m_D$.  The pion mass is comparable
with $\Delta^{(c)}$ and so the
relations in eqs. (14.3) and (14.4) break down for very small $v \cdot
p_\pi$.  For $v \cdot p_\pi \gg \Delta^{(c)}$ one recovers eq. (14.3)
from eq. (14.9) using the heavy quark flavor symmetry prediction for
the relation between $B$ and $D$ meson decay constants$^{[31, 32]}$
$$	f_B = \left[{\alpha_s (m_b)\over \alpha_s (m_c)}\right]^{-6/25}
\sqrt{{m_D\over m_B}} f_D \,\, . \eqno (14.11)$$

There are indications from lattice QCD,$^{[33]}$ QCD sum rules$^{[34]}$
and $1 + 1$ dimensional QCD$^{[35]}$ in
the large $N_c$ limit that the charm quark mass is not large enough for
the corrections to eq. (14.11) to be neglected.  Even if this is true,
eqs. (14.9) may still be a good approximation for both $B \rightarrow
\pi e\bar\nu_e$ and $D \rightarrow \pi \bar e \nu_e$.  If chiral $SU(3)_L
\times SU(3)_R$ is used then eqs. (14.9) can be used for $D \rightarrow
K\bar e\nu_e$ decay.  However, it is not clear that the kaon mass is
small enough for operators with one derivative to be neglected in
$L_\nu^3$.

\noindent {\bf 15.  Concluding Remarks}

These notes are meant to provide an introduction to chiral perturbation
theory for hadrons containing a single heavy quark.   For these hadrons
the combination of heavy quark and chiral symmetries is very powerful and
it makes a number of  interesting predictions.

Much of these notes (Chapters 2---11) consisted of a review of chiral
perturbation theory and heavy quark symmetry.  Applications of the
combination of these methods to properties of $B,B^*$ and $D,D^*$ mesons
were made in Chapters 11---14.  There has been considerable activity in
this area over the last year, and the few applications discussed in these
notes do not do justice to the breadth of applicability of the methods
developed here.  These notes provide the general background needed to do
research into the properties of heavy hadrons that can be studied with
chiral perturbation theory.  I encourage the reader to explore some of the
other applications of the combination of heavy quark and chiral symmetries
that have been discussed in the recently published literature.
\bigskip
\bigskip
\centerline {\bf References}
\item {1.} H. Georgi, Weak Interactions and Modern Particle Theory,
Benjamin/Cummings Publishing Co., Menlo Park, CA (1984); J. Donoghue, E.
Golowich and B. Holstein, Dynamics of the Standard Model, Cambridge
University Press (1992).
\item {2.} M. B. Wise, Particle Physics -- The Factory Era, Proceedings
of the Sixth Lake Louise Winter Institute, eds. B.A. Campbell, A.N.
Kamal, P. Kitching and F.C. Khanna, World Scientific (1991) 222.
\item {3.} H. Georgi, Proceedings of the Theoretical Advanced Study
Institute, eds. R. K. Ellis, C. T. Hill and J. D. Lykken, World
Scientific (1992) 589.
\item {4.} B. Grinstein, Annual Review of Nuclear Particle Science
{\bf 42} (1992) 10.
\item {5.} S. Weinberg, Physica (Utrecht) {\bf 96a} (1979) 327.
\item {6.} J. Bijnens, G. Ecker and J. Gasser, contribution to the
DAPHNE physics handbook, LNF--92--047--P (1992).
\item {7.} J. Wess and B. Zumino, Phys. Lett.  {\bf 37B} (1971)
95.
\item {8.} E. Witten, Nucl. Phys. {\bf B223} (1983) 422.
\item {9.} N. Isgur and M. B. Wise, Phys. Lett.  {\bf B232} (1989)
113; Phys. Lett.  {\bf B237} (1990) 507.
\item {10.} E. Eichten and B. Hill, Phys. Lett.  {\bf B234} (1990)
511;
\item {} H. Georgi, Phys. Lett. {\bf B240} (1990) 1446.
\item {11.} N. Isgur and M. B. Wise, Phys. Rev. Lett. {\bf 66}
(1991) 1130.
\item {12.} A. F. Falk and M. Luke, Phys. Lett. {\bf B292} (1992)
119.
\item {13.} E. Jenkins, A.V.  Manohar and M. B. Wise, Nucl. Phys. {\bf
B396} (1993) 27.
\item {14.} H. Georgi, Nucl. Phys.  {\bf B348} (1991) 293.
\item {15.} A. Falk, H. Georgi, B. Grinstein, \& M. B. Wise, Nucl.
Phys. {\bf B343} (1990) 1.
\item {16.} S. Nussinov and W. Wetzel, Phys. Rev. {\bf D36}
(1987) 1301.
\item {17.} M. A. Shifman and M. B. Voloshin, Sov. J. Nucl. Phys.
 {\bf 47} (1988) 199.
\item {18.} A. F. Falk, B. Grinstein and M. Luke, Nucl. Phys. {\bf B357}
(1991) 185.
\item {19.} M. Luke and A.V. Manohar, Phys. Lett {\bf B286} (1992) 348.
\item {20.} G. P. Lepage and B. A. Thacker, Nucl. Phys. {\bf B4} ( Proc.
Suppl.) (1988) 199.
\item {21.} M. B. Wise, Phys. Rev. {\bf D45} (1992) 2188.
\item {22.} G. Burdman and J. Donoghue, Phys. Lett. {\bf B208} (1992) 287.
\item {23.} T. M. Yan, H.Y. Cheng, C.Y. Cheung, G.L. Lin, Y.C. Lin and
H.L. Yu, Phys. Rev. {\bf D46} (1992) 1148.
\item {24.} S. Barlag et. al., (ACCMOR Collaboration) Phys. Lett. {\bf
B278} (1992) 480.
\item {25.} F. Butler et. al., (CLEO Collaboration)
Phys. Rev. Lett. {\bf 69} (1992) 2041.
\item {26.} J.F. Amundson, C.G. Boyd, E. Jenkins, M. Luke, A.V. Manohar,
J. Rosner, M. Savage, and M.B. Wise, Phys,. Lett. {\bf B296} (1992) 415.
\item {27.} P. Cho and H. Georgi, Phys. Lett. {\bf B296} (1992) 408.
\item {28.} L. Randall and M.B. Wise, Phys. Lett. {\bf B303} (1993) 139
\item {29.} N. Isgur and M.B. Wise, Phys. Rev. {\bf D42} (1990) 2388.
\item {30.}  N. Isgur and M.B. Wise, Phys. Rev. {\bf D41} (1990) 151.
\item {31.} M.B. Voloshin and M.A. Shifman, Sov. J. Nucl. Phys. {\bf 45}
(1987) 292.
\item {32.} H.D. Politzer and M.B. Wise, Phys. Lett. {\bf B206} (1988)
1681; Phys. Lett. {\bf B208} (1988) 1504.
\item {33.} A. Abada, et al., Nucl. Phys., {\bf B376} (1992) 172; C.
Bernard, C.M. Heard, J. Labrenz and A. Soni, Proc. Lattice 91, Tsukuba,
Japan (1991); S. Sharpe, Nucl. Phys. B. (Proc. Suppl.) {\bf 17} (1990)
146; A. Duncan E. Eichten, A. El-Khadra, J.M. Flynn, B. Hill and H.
Thacker, FERMILAB-Conf-92-331-T, Talk given at the International
Symposium on Lattice Field Theory; Lattice 92, Amsterdam, Netherlands
(1992).
\item {34.}  M. Neubert, Phys. Rev. {\bf D45} (1992) 245.
\item {35.}  B. Grinstein and P.F. Mende, Phys. Rev. Lett. {\bf 69}
(1992) 1018; M. Burkardt and E. Swanson, Phys. Rev. {\bf D46} (1992)
5083.
\vfil\eject
\centerline{{\bf Figure Captions}}

\item{{\rm Fig.~1.}} ~One loop Feynman diagram contributing to $\pi\pi
\rightarrow \pi\pi$ scattering.

\item{{\rm Fig.~2.}} ~Diagrams for the hadronic matrix element in $K^0
\rightarrow \pi^- \pi^0 e^+ \nu_e$ semileptonic ~decay.

\item{{\rm Fig.~3.}} ~Feynman diagrams that give nonanalytic $m_q^{1/2}$
contribution to the $D^* \rightarrow D\gamma$ ~matrix element.

\item{{\rm Fig.~4.}} ~Feynman diagram that give correction to heavy quark
symmetry predictions ~for $B\rightarrow D$ and $B\rightarrow D^*$ matrix
elements at zero recoil.

\item{{\rm Fig.~5.}} ~Tree graphs that determine $B \rightarrow \pi e
\bar\nu_e$ near zero recoil.

\bye